\documentclass[namedreferences]{solarphysics}
\usepackage{xcolor}
 
\usepackage[optionalrh,showbiblabels]{spr-sola-addons} 
\usepackage{graphicx}        
\usepackage{color}           
\usepackage[shortlabels]{enumitem}

\usepackage{hyperref}

\chardef\us=`\_

\usepackage[printonlyused,withpage]{acronym}
\begin{document}

\begin{article}
\begin{opening}

\title{A Machine Learning Enhanced Approach \\ for Automated Sunquake Detection \\in Acoustic Emission Maps}

\author[addressref={aff1},corref,email={mercea.vanessa@gmail.com}]{\inits{V.}\fnm{Vanessa}~\lnm{Mercea}}
\author[addressref=aff2]{\inits{A.R.}\fnm{Alin Razvan}~\lnm{Paraschiv}\orcid{0000-0002-3491-1983}}
\author[addressref=aff2]{\inits{D.A.}\fnm{Daniela Adriana}~\lnm{Lacatus}\orcid{0000-0003-2123-6605}}
\author[addressref={aff1},corref]{\inits{A.}\fnm{Anca}~\lnm{Marginean}\orcid{0000-0001-8426-588X}}
\author[addressref=aff3] {\inits{D.}\fnm{Diana}~\lnm{Besliu-Ionescu}\orcid{0000-0002-9864-0414}}

\address[id=aff1]{Technical University of Cluj-Napoca, Cluj-Napoca, 400114, Romania}
\address[id=aff2]{High Altitude Observatory, National Center for Atmospheric Research, PO box 3000, Boulder, Colorado 80307-3000, USA}
\address[id=aff3]{Astronomical Institute of the Romanian Academy, Str. Cutitul de Argint, Nr. 5, Bucharest, 040557, Romania}

\runningauthor{V. Mercea et al.}
\runningtitle{ML sunquakes}

\begin{abstract}

Sunquakes are seismic emissions visible on the solar surface, associated with some solar flares. Although discovered in 1998, they have only recently become a more commonly detected phenomenon. Despite the availability of several manual detection guidelines, to our knowledge, the astrophysical data
produced for sunquakes is new to the field of Machine Learning. 
Detecting sunquakes is a daunting task for human operators and this work aims to ease and, if possible, to improve their detection. 
Thus, we introduce a dataset constructed from acoustic egression-power maps of solar active regions obtained for Solar Cycles 23 and 24 using the holography method. We then present a pedagogical approach to the application of machine learning representation methods for sunquake detection using AutoEncoders, Contrastive Learning, Object Detection and recurrent techniques, which we enhance by introducing several custom domain-specific data augmentation transformations. We address the main challenges of the automated sunquake detection task, namely the very high noise patterns in and outside the active region shadow and the extreme class imbalance given by the limited number of frames that present sunquake signatures. 
With our trained models, we find temporal and spatial locations of peculiar acoustic emission and qualitatively associate them to eruptive and high energy emission. While noting that these models are still in a prototype stage and there is much room for improvement in metrics and bias levels, we hypothesize that their agreement on example use cases has the potential to enable detection of weak solar acoustic manifestations.
\end{abstract}

\end{opening}

\section{Introduction}
     \label{S-Introduction} 

\indent \indent
Local helioseismology has been the primary tool used to study occasional emission of seismic transients, more commonly known as sunquakes, coming from the solar surface or from sources submerged in the  solar interior that are sometimes released by solar flares. \citet{kosovichev98} first discovered sunquakes as expanding rings in the Dopplergram data using a factor of four image-enhancement technique. The technique revealed the expanding rings of an almost circular-shaped surface ripple after the 9 July 1996 X2.6 solar flare. \citet{doneaetal99} obtained egression-power maps using the method developed by \citet{lindseybraun00} for this particular sunquake event. Since then, sunquakes have become a more commonly detected phenomenon  \citep{donea2011,kosovichev2011, besliu-ionescuetal2017,sharykin2020}. Typically, sunquakes are associated with intense reconnection events in the solar atmosphere resulting in strong solar flares, of X- or M- spectral class, although \citet{Sharykin+2015} found signatures related to a weaker C.7 class event.

There are several methods to detect a seismic emission. 
From a chronological point of view these are: time--distance analysis \citep{kosovichev98}, seismic holography \citep{lindseybraun00}, and  seismic ripples detection (movie method) \citep{kosovichev98,kosovichev2011,sharykin2020}.
Each method has its advantages and disadvantages; some of them show the source, some the wave propagation, and usually not all methods show signatures for the same flares \citep{sharykin2020}.

Sunquake selection criteria require a continuous detection of acoustic signal of at least a couple of minutes and a reasonable source signal-to-noise ratio. These criteria are further extended in Section \ref{Ss-method-dt}. More elaborate and precise selection criteria have been proposed in the literature \citep[e.g.][]{chen2021}, where a sunquake selection rule can be the conjuncture between impulsive flaring events and downward background oscillatory velocity, occurring at the same location. In practice, an intricate analysis needs to be done on a flare-by-flare basis on the entire dataset. We thus could not utilize such criteria in this work.

Despite the variety of available methods for sunquake detection, an attempt to automate the detection process in order to reduce human effort has yet to emerge. To facilitate this, we construct a Machine Learning (ML) ready dataset covering SC23 and SC24.
To our knowledge, these specialized astrophysical data products were not previously processed via ML techniques. This work takes the first steps in this direction and describes the components of several ML-enhanced detection models, as a result of an extensive array of experiments centered on Representation Learning techniques. See \citet {bengio2013representation} for a review of these techniques. 
We discuss findings from our experiments and identify challenging aspects. 

This work is structured as follows: Section \ref{S-method} describes the methods used for data ingestion, pre-processing, holography analysis, and preparation for ML algorithm ingestion. The ML methodology is described in detail in Section \ref{Ss-method-ml}, in preparation for Section \ref{s-results}, where particularities and limitations of two detection models are presented along with an analysis of current results on both known and tentative newly-detected sunquake events. Lastly, Section \ref{S-Conclusion} summarizes the discussion of models limitations and main results, and suggests avenues of interest for future development. Furthermore, in Appendix \ref{sss-autoencoder}\,-\,\ref{od-appendix} we expand on additional methods and more standard methodological approaches that have been explored while pursuing the task at hand that have not been included in the final proposed solution.

\section{Data Ingestion and Pre-Processing} 
      \label{S-method}    
\subsection{The Helioseismic Holography Method}    \label{Ss-method-dt}  

In this work, we have used the helioseismic holography analysis of \citet{lindseybraun00} to process raw data and identify the seismic emission during solar flares from a selected list of sunquake events. The method is applied to photospheric Dopplergram maps such as those provided by the \textit{Michelson Doppler Imager} \citep[MDI:][]{scherrer1995} onboard the \textit{Solar and Heliospheric Observatory} \citep[SOHO:][]{soho1995} and \textit{Helioseismic and Magnetic Imager} \citep[HMI:][]{sdohmi2012} onboard the \textit{Solar Dynamics Observatory}  \citep[SDO:][]{sdo2012}. 

In general terms, the helioseismic-holography technique is used to image acoustic sources on and beneath the Sun's photosphere. It reconstructs phase-coherent $p$-mode acoustic waves that are observed at the photosphere into the solar interior to render stigmatic images of the subsurface sources that have perturbed this surface.

The solar interior refracts downward-going waves back
toward the Sun's surface due to the temperature gradient below the photosphere leading to increasing sound speed towards the interior. Helioseismic holography images a selected area, in a way that is “broadly analogous to how the eye treats electromagnetic radiation at the surface of
the cornea, wave-mechanically refocusing radiation
from submerged sources to render stigmatic images”
\citep[][]{lindseybraun00}. In order to obtain these
images, holography uses a pupil defined as an annulus
with radius 15\,--\,45\,Mm, to image the focus situated a
considerable distance from the pupil and computes the “ingression”, [H$_-$], and “egression”, [H$_+$]. 
The ingression, and the egression, are obtained from the wave-field at the surface, $[\Psi]$, through theoretical Green’s functions \citep[][]{lindseybraun00}.

The egression power P(\textsl{\bf r},{\it t})=$|$H$_+$(\textsl{\bf r},{\it t})$|^2$ is extensively used in detecting or studying acoustic sources and absorbers \citep{dianathesis}. This equation is used to calculate the egression power for each pixel in the image. Therefore,
using this technique, we can create maps of egression
power around ARs in order to detect the seismic sources.

Following \citet{dianathesis}, we employ a temporal selection criteria in which a eight-minute duration of the egression power signature is required for a positive sunquake identification due to an artifact of the truncation of the helioseismic spectrum by a 2\,mHz pass-band. The egression power signatures that result are temporally smeared to a minimum effective duration of order 

\begin{equation}
\Delta t ~=~ \frac{1}{\Delta \nu}
~=~ \frac{1}{2~{\rm\,mHz}}
~\approx~ 500~{\rm\,seconds}.
\label{eqdt}
\end{equation}

Other selection criteria using this technique concern the position of the source as relative to the active region (AR), its intensity in terms of a 3$\sigma$ pixel enhancement, and a clear increase--decrease type signal, or even more complex criteria as suggested by \citet{chen2021}. For this ML-oriented work, where absolute intensity measurements tend to lose significance, the most important applicable constraint is the temporal selection criteria. 

The band centered at 6\,mHz  was chosen for this work because it showed a much higher signal-to-noise ratio compared to other bands \citep{doneaetal99} making it easier to detect the source \citep[egression power in absolute terms $\approx4.1$ times that of the 6\,mHz mean quiet Sun,][]{doneaetal99}.

\subsection{Sunquake Identification and Dataset Creation}    \label{Ss-method-sl} 

For this work, we produced a sunquake dataset spanning two solar cycles, SC23 and SC24, using the holography method described above. To identify potential sunquakes candidates, we have used the curated lists of \citet{dianasq23_2012} for SC23, totaling 15 MDI observed events, along with the positive holography identifications from SC24 of \citet{sharykin2020} totaling 80 events. In the case of SC24, multiple sunquakes have been identified in the same AR, but at different times during its disk crossing. For the complete list and parameters of sunquakes that were positively identified and marked ``+" by \citet{sharykin2020} (see their Table 1).
To create our dataset, we used temporal cubes of 5\,--\,7\,mHz holography-acoustic-egression maps binned to resolutions of $256\times256$ pixels that lead to approximately 1.5$^{\prime\prime}$ in physical resolution 
leading

to $\approx200\times200^{\prime\prime}$ acoustic maps. This binning is required in order to get adequate signal-to-noise ratios necessary for a positive sunquake detection. For all positive events in both SC23 and SC24, we have created acoustic datacubes spanning three-hour intervals, as required by the holography methodology, with observations starting approximately 1\,--\,1.5 hours before the sunquake inducing flare, and following for an additional 1.5\,--\,2 hours. The data cadence of MDI (SC23) is 60 seconds, while the data cadence of HMI (SC24) is 45 seconds. The before and after temporal limits are flexible in order to not bias an ML analysis by generating negative and positive frames at the same temporal location in each individual cube. This setup generates severely imbalanced datasets, where only 3\,--\,4\,\% of the observations contain expected sunquake signal. Small datagaps of less than eight frames ($<$360\,--\,480 seconds $<$ minimum effective duration) have been deemed safe to be interpolated in the cubes.  

The very high noise patterns outside ARs along with oscillatory moving patterns that result from the holographic analysis (see Figure \ref{fig:samples.png}) make a very challenging dataset. Because of this, we have not considered the number of sunquakes in the two input lists to be definitive. Our curated positive events for our ML-model training represent only the clearly identifiable and distinguishable sunquakes that were inspected and annotated by us based on the sunquake detection criteria discussed in Section \ref{Ss-method-dt}. Adding to this, the holography analysis on solar ARs is generally considered a convoluted process that requires manual input. We have bypassed these limitations by creating a Sunpy Version 4.0.x \citep{sunpy_community2020, stuart_j_mumford_2022_6974892} enabled batch script where we have queried and downloaded magnetogram and dopplergram data from the MDI and HMI repositories corresponding to the temporal slots around our selected sunquakes. We then computed the central heliographic Stonyhurst to Carrington position of the AR at the time of the sunquakes by retrieving and interpolating data from the solarmonitor service (\href{www.solarmonitor.org}{\textsf{www.solarmonitor.org}}). The positions were automatically ingested into the holography method along with a set of other fixed parameters, enabling us to process the sunquake lists and produce temporal cubes of acoustic egression power corresponding to each event.   

The final acoustic datasets from both SC23 and SC24 that are employed by the methodologies described below, are available in the following online Kaggle SunquakeNet repository \citep{kagglerepo}.
The repository also includes the region selections resulting from manual identifications of sunquakes from \citet{dianasq23_2012} and \citet{sharykin2020}. Individual frame datasets ready for ML ingestion in JPEG format separated into positive vs. negative sunquake detections spanning all utilized sunquake datacubes are also included.

\subsection{ML Dataset Creation}

\label{Sss-data-ml}
Before diving into the ML-driven methodology, we will first describe the processes that data undergoes in preparation for model ingestion. 

\begin{figure}[ht]
\centering
\includegraphics[width=\textwidth]{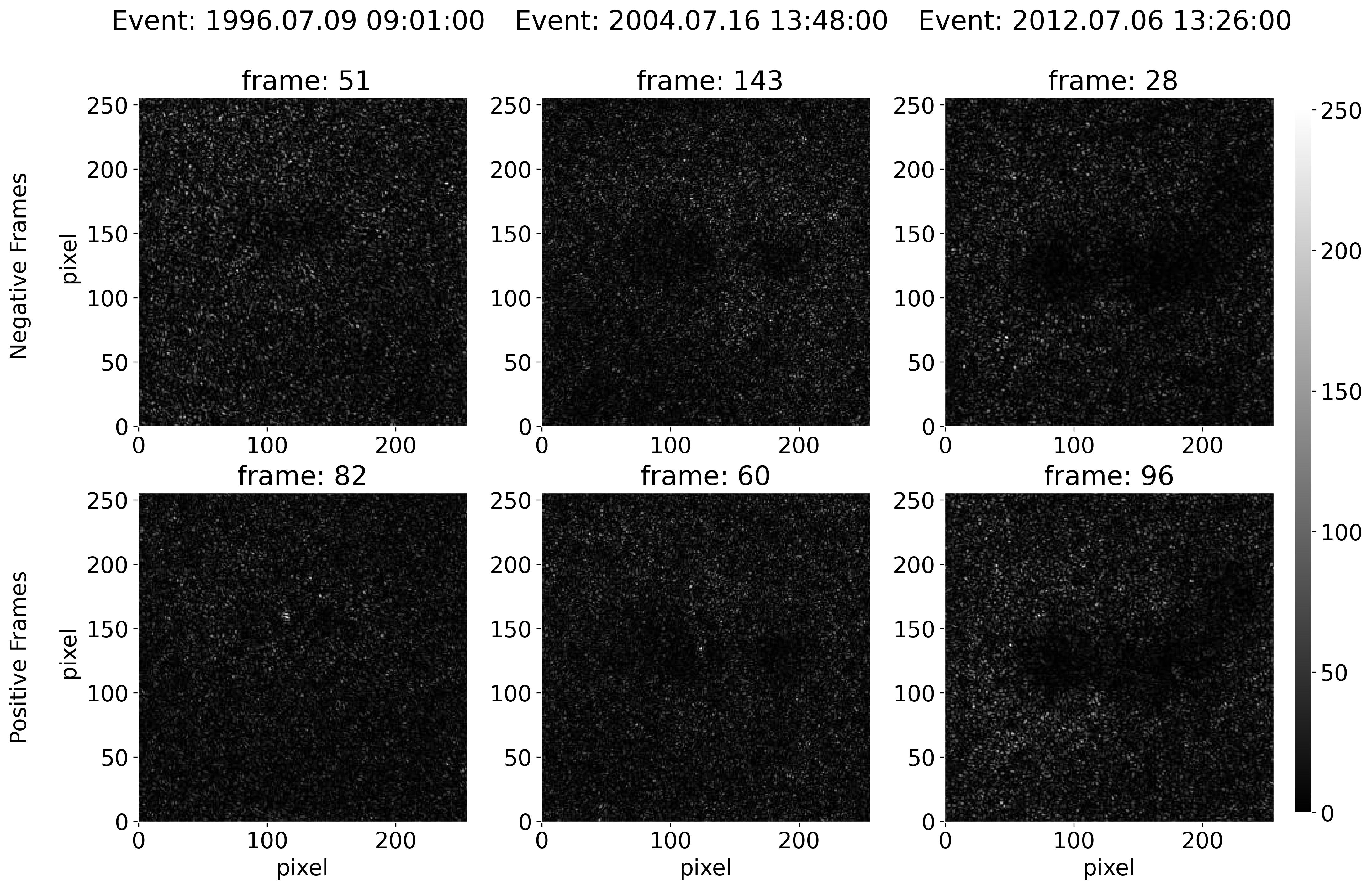}
\caption{A compilation of six 8-bit scaled intensity  data samples from the ML-prepared dataset. Row titles indicate the label and column titles indicate event identifiers and frame indices.}
\label{fig:samples.png}
\end{figure}

To create an ML-suited dataset based on the egression-power maps, frames are extracted from the volumes into gray-scale $224\times224$ pixels PNG files and a MinMax normalization is applied to each datacube. The obtained images are labeled as follows. An image is considered to be positive, i.e. to contain a sunquake, if the corresponding frame in the egression power map cube shows the presence of a sunquake, and negative otherwise. To identify whether a frame is positive, the starting time present in the egression power map file header is used to describe a point in time and respectively, a corresponding frame range for one sunquake. Based on this equivalence, the sunquake begin and end times are correlated to frame indices. Figure  \ref{fig:samples.png} depicts six randomly extracted negative and positive samples from the ML dataset.

One reason for dividing the volumes into 2D images lies in the small quantity of event samples at hand, with only 15 volumes containing sunquakes recorded in SC23, and 38 volumes containing sunquakes in SC24 annotated at the time of this research. An additional 17 datacubes with no events are available from SC23 but are not used in the proposed models due to already increased class imbalance. Moreover, 43 additional cubes from SC24 have been downloaded and processed, but they present weaker and less-clear signatures, caused by the noise threshold inside the ARs achieved via holography. In these cases the manual sunquake identifications were less certain, where in few cases we are unsure if recovering any helioseismic signal at all. As shown by \citet{sharykin2020}, different events were sensible to different methods, where here we only employed holography. Thus, these events were dropped from the analysis as our main scope revolves around the correct identification of clear seismic signatures. Although this increases the imbalance, not including these events in the training and validations of current preliminary models will ensure we are not inducing incorrect information when testing different ML approaches. In a subsequent project, including these less certain events will be precisely the focus for fine-tuning the models, after identifying a suitable threshold for applying holography.

By dividing the volumes, the sequence structure of the data is not disregarded, as a significant part of the experiments, including the models in Section \ref{s-results}, 
introduce a form of sequence modeling. A second reason behind this division is the wider variety of deep learning methods available for 2D-image datasets as opposed to movielike datasets.  

From the sunquake catalog presented in Table \ref{table:used_sunquake_events}, 53 of the listed events are used for model training and validation, marked with ``+". An additional four datacubes without annotated sunquake signatures are used for testing and analysis, marked with ``-". These latter sets are used to analyze the results for the proposed approaches, with the goal of identifying emissions that are too weak to be detected using conventional methods. 

The level of imbalance in the obtained ML dataset is a challenging aspect. By combining both SC23 and SC24 data, the quantity of positive images in the dataset totals 845 (205 + 640) while the negative count totals 13,055 (3891 + 9164). Some ``blank" all-black buffer frames in datacubes are of course excluded from the counts and analysis.

\begin{table}
\tiny
\begin{tabular}{c c c c}
Event & SQ Start Frame Index & SQ End Frame Index & Labeled \\\hline
09 July 1996 09:01	&	69	& 87 & + \\ 
06 June 2000 14:57	&	47	& 66 & + \\
24 November  2000 04:54 	&	62	& 76 & + \\
06 April 2001 19:13	&	21	& 36 & + \\
10 April 2001 05:01	&	47	& 56 & + \\
24 September 2001 09:35	&	38	& 48 & + \\
15 July 2002 19:52 	&	60	& 74 & + \\
23 July 2002 00:27	&	30	& 43 & + \\
21 August 2002 05:24	&	43	& 58 & + \\
23 October 2003 08:45	&	34	& 42 & + \\
28 October 2003 11:00	&	170	& 178 & +  \\
29 October 2003 20:35	&	138	& 149 & +  \\
16 July 2004 13:48	&	56	& 69 & + \\
13 August 2004 18:07	&	15	& 24 & + \\
15 January 2005 00:33	&	40	& 54 & + \\
15 February 2011 01:44	&	92	& 104 & + \\
18 February 2011 09:55	&	107	& 117 & + \\
18 February 2011 12:59	&	103	& 111 & + \\
07 September 2011 22:32	&	82	& 105 & + \\
24 September 2011 20:29	&	90	& 99 & + \\
25 September 2011 08:46	&	-	& - & - \\
26 September 2011 05:06	&	129	& 139 & + \\
30 December 2011 03:03	&	-	& - & - \\
05 March 2012 19:27	&	89	& 109 & + \\
06 March 2012 04:01	&	95	& 110 & + \\
06 March 2012 07:52	&	82	& 94 & + \\
06 March 2012 12:23	&	106	& 122 & + \\
09 March 2012 03:22	&	88	& 103 & + \\
08 May 2012 13:02	&	-	& - & - \\
09 May 2012 21:01	&	93	& 114 & + \\
10 May 2012 04:01	&	-	& - & - \\
04 July 2012 09:47	&	112	& 130 & + \\
04 July 2012 14:35	&	88	& 103 & + \\
05 July 2012 03:25	&	92	& 106 & + \\
05 July 2012 11:39	&	89	& 104 & + \\
05 July 2012 20:09	&	87	& 106 & + \\
06 July 2012 01:37	&	82	& 96 & + \\
06 July 2012 13:26	&	84	& 100 & + \\
24 October 2013 10:30	&	89	& 98 & + \\
06 November  2013 13:39	&	86	& 103 & + \\
07 November  2013 03:34	&	84	& 105 & + \\
08 November  2013 04:20	&	85	& 104 & + \\
10 November  2013 05:08	&	87	& 103 & + \\
02 February 2014 06:24	&	SQ1: 90, SQ2: 105	& SQ1: 108, SQ2: 114 & + \\
07 February 2014 10:25	&	86	& 120 & + \\
09 November  2014 15:24	&	88	& 102 & + \\
03 January 2015 09:40	&	89	& 106 & + \\
10 March 2015 23:46	&	95	& 109 & + \\
11 March 2015 16:11	&	90	& 105 & + \\
22 August 2015 21:19	&	89	& 97 & + \\
28 September 2015 14:53	&	85	& 104 & + \\
30 September 2015 13:18	&	85	& 97 & + \\
04 September 2017 20:28	&	86	& 102 & + \\
05 September 2017 01:03	&	86	& 100 & + \\
06 September 2017 08:57	&	91	& 111 & + \\
07 September 2017 10:11	&	86	& 101 & + \\
07 September 2017 14:20	&	99	& 115 & + \\
\hline
\end{tabular}

\caption{Sunquake events between 09 July 1996 and 15 January 2005 from SC23 and between 15 February 2011 and 07 September 2017 from SC24. Events marked with ``+" are used in training and validating the proposed Representation Learning models. Events marked with ``-" are part of an additional dataset, and not sunquake annotated, and are used for post-learning testing.}
\label{table:used_sunquake_events}
\end{table}

\section{Machine Learning Approaches for Sunquakes}    \label{Ss-method-ml}  

To derive a detection model able to capture sunquake signatures, several experiments are performed using ML techniques of increasing complexity. These can be divided into two phases: Methods described in Appendix \ref{sss-autoencoder} and Section \ref{sss-objectdet} are trained on SC23 data; Methods in Section \ref{sss-contrastive} are trained using the combined SC23 and SC24 datasets, based on the process presented above (Section \ref{Ss-method-dt}). The decision to combine the datasets is based on preliminary findings, which indicated that given the low data regime, the amount of AR morphologies that an ML model is exposed to needs to be increased such that the model is able to shift the focus from learning representations of ARs to learning sunquake signatures.

The main ML metrics that will be used to judge model performance are: Precision ($\frac{TP}{TP+FP}$), Recall ($\frac{TP}{TP+FN}$), F1-Score ($\frac{2\,Precision\,Recall}{Precision+Recall}$) and Accuracy ($\frac{TP+TN}{TP+FP+TN+FN}$). The $TP$ and $FN$ labels denote true and false positives, and $TN$, $FN$ the true and false negatives. $GT$ will be used to denote Ground Truth, or real label.

Because the data has domain specific particularities, learning useful representations from the input has been a primary research goal. Initial experiments are centered on less complex methodologies, starting with small scale Convolutional Neural Networks (CNNs), which prove insufficient. Transfer Learning from ImageNet \citep{deng2009imagenet} with common CNN architectures also failed to converge to fully reliable performance metrics. One possible explanation is that ImageNet contains natural images that are quite different from our data, and consequently, the features captured by the pre-trained networks are not sufficiently relevant for our task characterised also by a limited and imbalanced dataset. For this reason, we decided to first move towards Representation Learning in the form of AutoEncoder approaches, which also rendered less-satisfactory performance. These are described in more detail in Appendix \ref{sss-autoencoder}. 

In this paper, we focus on Contrastive Learning (CL) methods and results, covering supervised (that rely on annotations) and unsupervised (that do not use annotations) objectives, recurrence techniques and observations on relevant data augmentations. The next subsections describe these experiments and methods, and highlight the identified challenges and limitations. 

\subsection{Contrastive Learning Approach}
\label{sss-contrastive}

Experiments performed on the holography data indicated that the main challenges in sunquake classification include class imbalance, low data regime problems, and the inability of AutoEncoder-based approaches to capture the relevant sunquake features in the latent distribution. As a result, the second phase of the experiments is focused on a more recent computer-vision methodology, namely the CL. \citet{CAO202171} argue that, as opposed to AutoEncoder methods, the goal of CL is not that of finding exact distributions for data samples, but rather about discriminating between different samples.

CL initially emerged as a self-supervised method for learning from visual representations and showed relevant improvements over previous state-of-the-art models, matching supervised network performances \citep{pmlr-v119-chen20j}. The goal of CL is intuitive: train a network to generate close latent space representations for pairs of data points that are similar to one another and distinctive representations for dissimilar pairs.

\begin{figure}[t]
    \begin{minipage}[t]{0.48\linewidth}
        \includegraphics[width=1\textwidth]{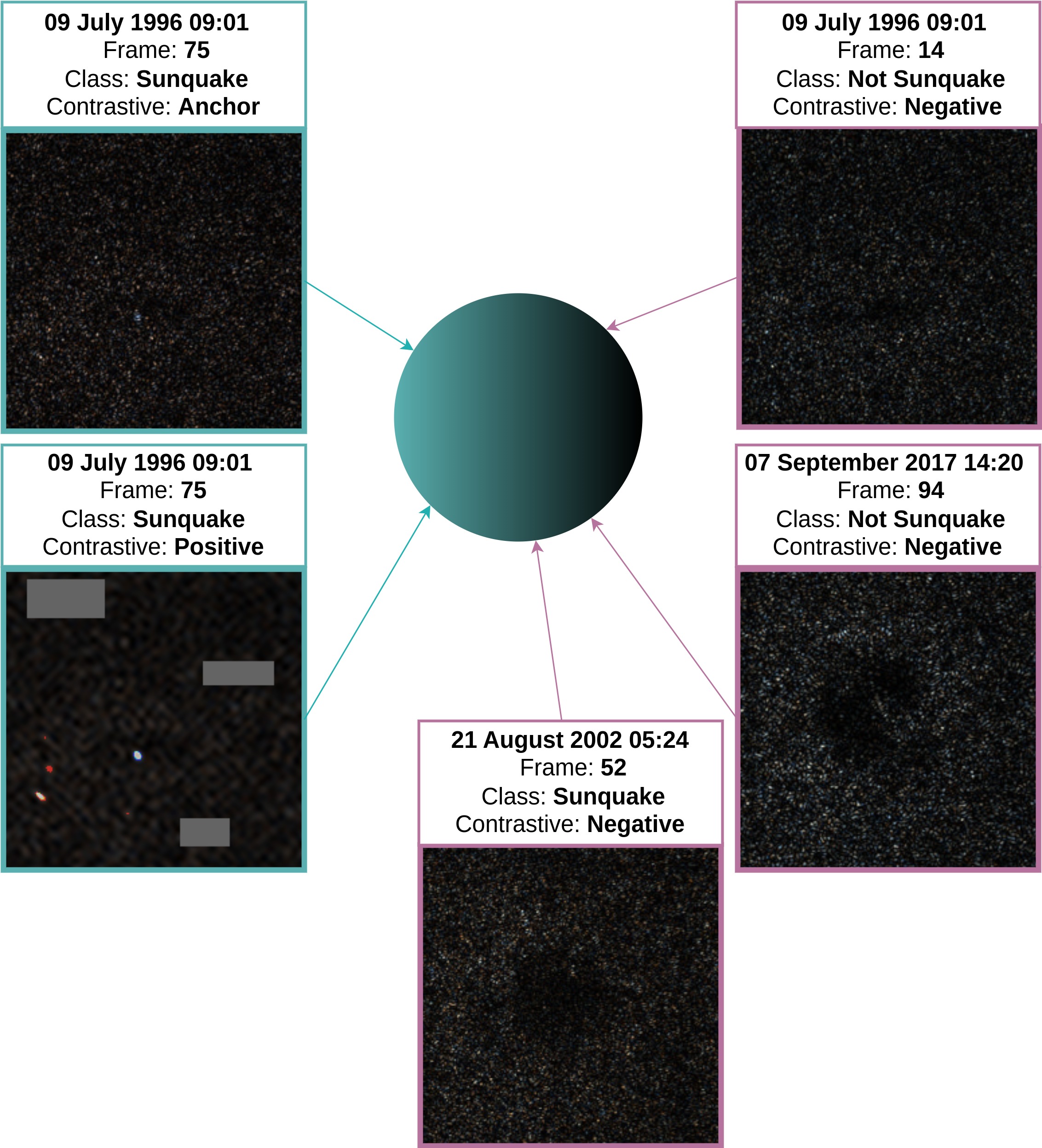}
    \end{minipage}
    \hfill \vline \hfill \hfill
    \begin{minipage}[t]{0.48\linewidth}
        \includegraphics[width=1\textwidth]{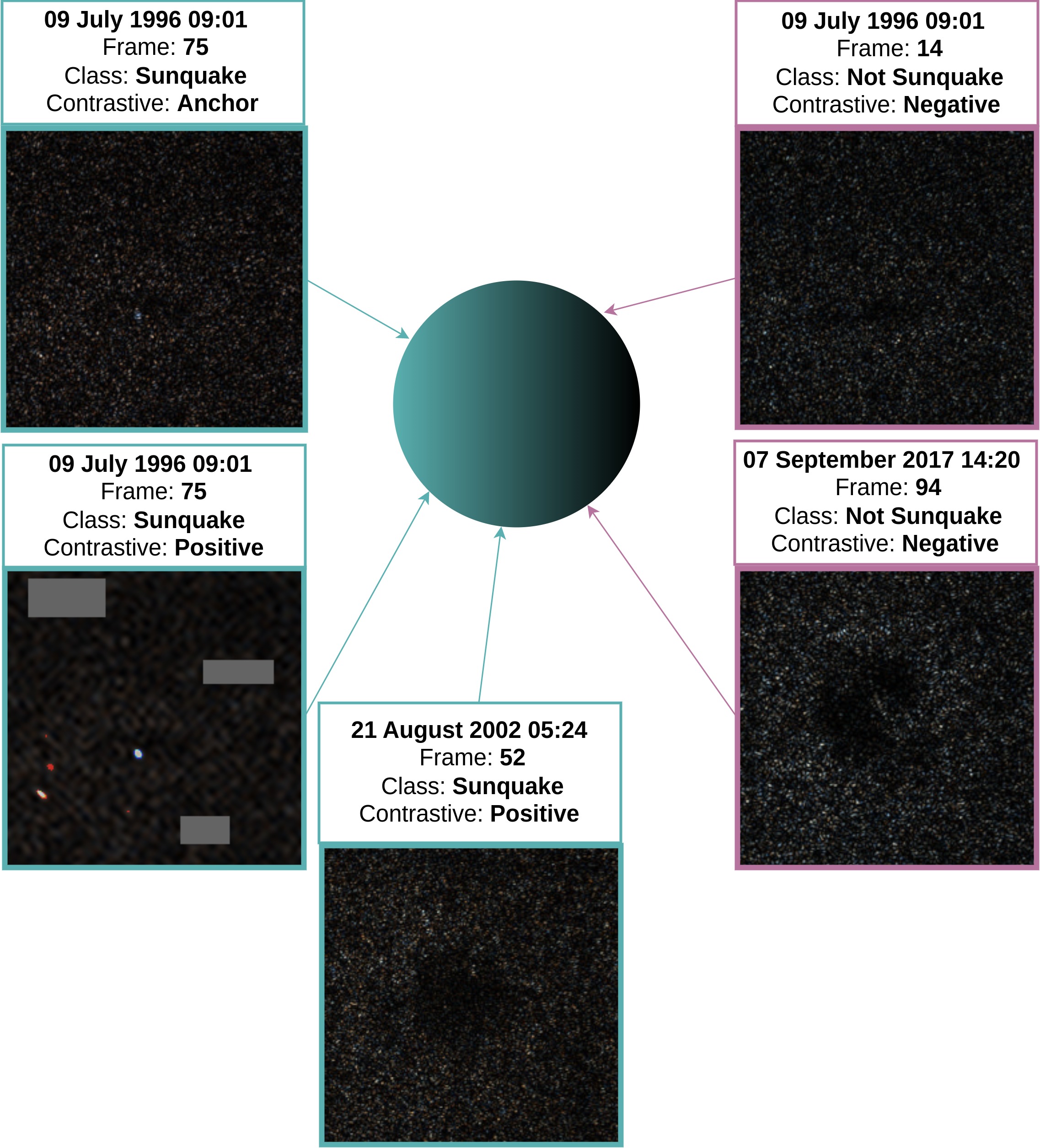}
    \hfill
    \end{minipage}
\centering
\caption{Self-Supervised \textit{(left)} and Supervised Contrastive Learning \textit{(right)} Visualization for sunquake image datasets. The self-supervised contrastive loss contrasts the anchor and an augmented version of it (positive) against all other samples (negative), regardless of class. The Supervised contrastive loss contrasts the anchor, an augmented version of it, and all the data samples of the same class (positives) against all other samples (negatives).}
\label{fig:self_supervised_contrastive}
\end{figure}

Recently, a supervised approach to CL has been proposed by \citet{supcon}, where clusters of points belonging to the same class are pulled together in the embedding space, while simultaneously pushing apart clusters of samples from different classes. Figure \ref{fig:self_supervised_contrastive} offers an overview of the differences between self-supervised and supervised CL approaches.

The literature presents a number of contrastive loss functions: max margin loss  \citep{max_margin_loss}, triplet loss \citep{triplet_loss}, multi-class N-pair loss \citep{npair_loss}, SimCLR loss \citep{pmlr-v119-chen20j}, but the main idea behind this method of learning has not changed drastically over the years. 

Appendix \ref{sss-self-sup-cl} describes experiments where a fully self-supervised approach is pursued with the goal of avoiding the problems arising from class imbalance. On average, results of this approach are similar to those of the AutoEncoder methods. We note that this approach is not able to capture many relevant features. Appendix \ref{sss-sup-cl} describes experiments with a supervised-contrastive model, where we find that some relevant and sometimes distinct features are captured. Metric improvements over the self-supervised model are encouraging but the class imbalance factor still posed a great impact on the classification results. 

To tackle this, we combine the Self-Supervised and Supervised contrastive models in a unitary pipeline using a Two-Step approach. Self-supervised and Supervised contrastive loss functions are shown in Equations \ref{eq1}-\ref{eq-supcon} and are described in detail in Appendices \ref{sss-self-sup-cl} and \ref{sss-sup-cl} respectively. 
For the self-supervised contrastive training, positives are upsampled with geometric transforms. 
For the supervised part, class weighting is introduced in the contrastive loss, consistent to the equations of \citet{zhong2022rescom} in order to tackle the imbalance effect.
Under this loss, where the tailed class samples are assigned larger weights, the model achieves the highest precision.
The updated weighted supervised contrastive loss is:
\begin{equation}
\label{eq2}
l_{i} = \sum_{i \in I} \frac{-w_{y_i}}{|P(i)|} \sum_{p \in P(i)} \textrm{log} \frac{\textrm{e}^\frac{\textrm{sim}(z_{i} z_{p})}{\tau}}{\sum_{k=1}^{2N} \textrm{mask}_{[k\neq i]}\textrm{e}^\frac{\textrm{sim}(z_{i} z_{k})}{\tau}} \; ,
\end{equation} 
\begin{equation}
w_{y_{i}} = \frac{1 - \beta}{1 - \beta ^ {N_{y_{i}}}}.
\end{equation} 
where $\beta \in [0, 1)$ is a hyper-parameter and $\frac{1}{w_{y_{i}}}$ is the effective number of samples for class $y_i$. In our work, a value of $0.9999$ is used for $\beta$ to halve the effective number of negative samples. During training, Adam \citep{kingma2014adam} and Stochastic Gradient Descent (SGD) 
are used as optimizers. The prior is quickly diverging for the majority of experiments. SGD with momentum, weight decay, and warm-up is found to generalize better.

ResNet-18, Resnet-50, and DenseNet-121 \citep{densenet} backbones are considered case by case, in an attempt to improve results by increasing model complexity. 

We note that a set of augmentations additionally play an important role in convergence. These are discussed below. Two relevant models extracted from the above experiments and their results are analyzed in depth in Section \ref{s-results}.

\subsection{Data Augmentation and Sampling}
\label{sss-augmentations}    

To lower the impact of the class imbalance, we experiment with different transforms for augmenting our input data, including: center and random crops, sharpness, color jitter, posterize, invert, auto contrast, solarization, Gaussian blur, normalization, vertical/horizontal flips, and random rotations of 90\textdegree, 180\textdegree, 270\textdegree. In Appendix \ref{sss-augmentations2} we detail the main transforms that are applicable, following the categories of transformations described by \citet{augmentations2022Yang}. We comment on the applicability of each of these augmentation approaches as not all of these usually standard transforms are useful for our particular task, some being detrimental to the learning process. We iterate below only custom transforms, that are applied to our sunquake data in addition to the more generic approaches illustrated in Appendix \ref{sss-augmentations2}: 

\begin{enumerate}[i)]
    \item \textit{Custom Time-Based Mixing:} When experimenting with various time preserving techniques, we introduce an augmentation that combines subsequent gray-scale frames into a single '3D-like' sample. Channels are represented by consecutive frames. This technique enables the contrastive model to better focus on similarities between samples that are close to one another in both time and space. We remind the reader that sunquakes occur in successive frame series.\\

    \item \textit{Custom Solarized Low Pass Filter:} This method is a custom implementation of a color space transform (Appendix \ref{ssss-augmentations2-colorspace}). Given that the process of obtaining seismograms (maps of distances traveled by the wave front) includes a last step of applying a Fourier transform with respect to the azimuthal angle \citep{kosovichev98}, we believe that an augmentation based on such a transform may be useful in enhancing the sunquake details. We apply a low pass filter followed by a solarize transform. The threshold for both transforms is set to 50. We report that this mix of transforms enhances high-frequency signals, so that  sunquake features are amplified. This enhancement is shown in Figure  \ref{fig:solarized_low_pass_filter}, where a more intense sunquake spot is observed on the augmented images. One downside to this transform is that it also enhances high-frequency areas that are not necessarily sunquakes. A more evident example for this is shown in the third image in Figure \ref{fig:solarized_low_pass_filter}. However, when using this transform with a 50\,\% probability, the contrastive loss no longer stagnates during training, as it does with other transforms typically used in CL. While this statement applies to our data, we can not draw general conclusions about its applicability. For example, the use of this custom transform with the \textsf{CIFAR-10} dataset revealed a significant deterioration in the classification results (of $\approx$ 20\,\% in F1-Score). Because we use this augmentation to better infer domain knowledge to the CL approaches, we experimented with its limited applicability as a method to explain the CL predictions in Section \ref{ss-falsepos}. 

\begin{figure}[ht]
\includegraphics[width=\textwidth]{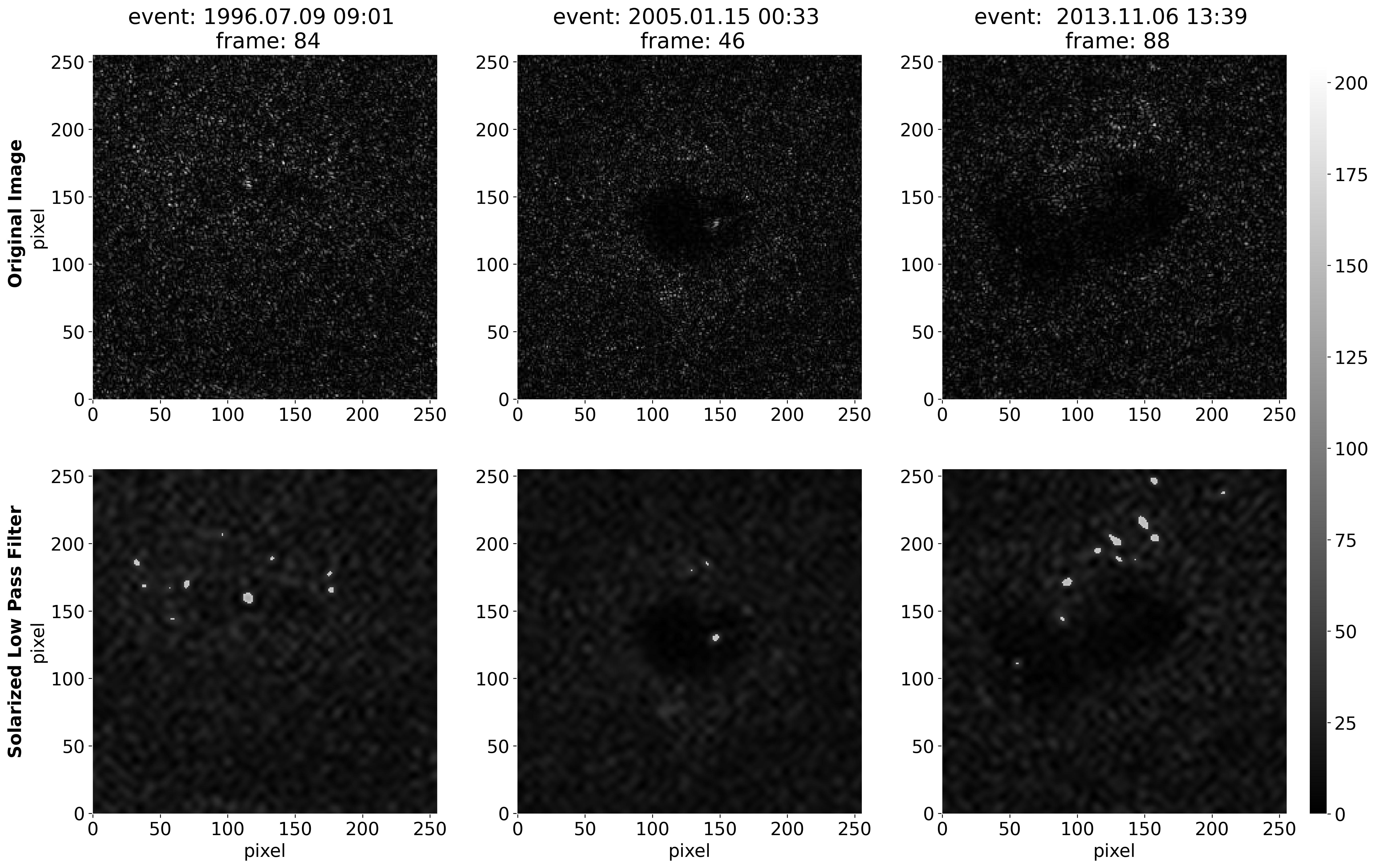}
\caption{Solarized low-pass filter applied to a selection of 8-bit intensity scaled frames from the SC23 and SC24 dataset. Frame indices and events are mentioned in the column titles. } 
\label{fig:solarized_low_pass_filter}
\end{figure}

    \item \textit{Customized Random Erasing:} Random erasing is detailed in Appendix \ref{ssss-augmentations2-occlusion}. Here, we develop and apply a custom Random Erasing implementation in order to provide a higher degree of label-preservation for our data. For this, we propose a method for decreasing the probability of occluding a sunquake: instead of adding standard large-sized rectangle(s) of erased areas, we augment with using multiple small erasing rectangles, of sizes covering up to 4\,\% of the original input image, with an up to 50\,\% probability of application. The smaller sized rectangles have a high probability of being applied, while the large sized ones have a significantly lower probability. For instance, 50\,\% of the images have an erasing rectangle covering 1\,\% of the image. 5\,\% of the images have an erasing rectangle covering 4\,\% of the image. Between zero and eight erasing regions may be applied to a singular image (see the \textit{Contrastive: Positive} image in Figure \ref{fig:self_supervised_contrastive}).

\end{enumerate}

\subsection{Recurrent Techniques}
\label{sss-reccurret}

During the dataset-preparation process described in Section \ref{Sss-data-ml}, the sequence component of the input data is lost. We hypothesize that the main reason for the modest results of single-frame approaches such as those of AutoEncoder-based models (Appendix \ref{sss-autoencoder}) lies in the inability of the models to capture sunquake signatures from only single data samples. Therefore, separate methods need to be introduced to maintain the time-series structure of the data, as sunquake signatures are only valid if visible at successive times and frames.

To reintroduce the sequence information, the output from the representation model is taken and then combined with previous and next sample embeddings using a sliding window of various temporal sizes. To interpret the clustering of the encoded data points dimensionality reduction techniques are applied, specifically Principal Component Analysis \citep[PCA:][]{MACKIEWICZ1993303} and Uniform Manifold Approximation and Projection \citep[UMAP:][]{umap}. 
The sliding window methodology is enhanced with the inclusion of the UMAP components for each selected window. This approach, when applied to Contrastive representations, yielded weak results (Macro Avg. Metrics: Precision 0.50, Recall 0.51, F1-Score 0.48, Accuracy 0.74), compared to its successor.

These more standard approaches still prove insufficient for contrastive representations. Losing the sequence information during image-based learning makes it difficult for the model to correlate consecutive samples of the same sunquake. Not all single frames that should be positively identified are recovered, as confirmed by the predictions. As an answer to this, we propose the custom Time Based Mixing transform described in Section \ref{sss-augmentations}.  

\begin{figure}[!t]
\includegraphics[width=\textwidth]{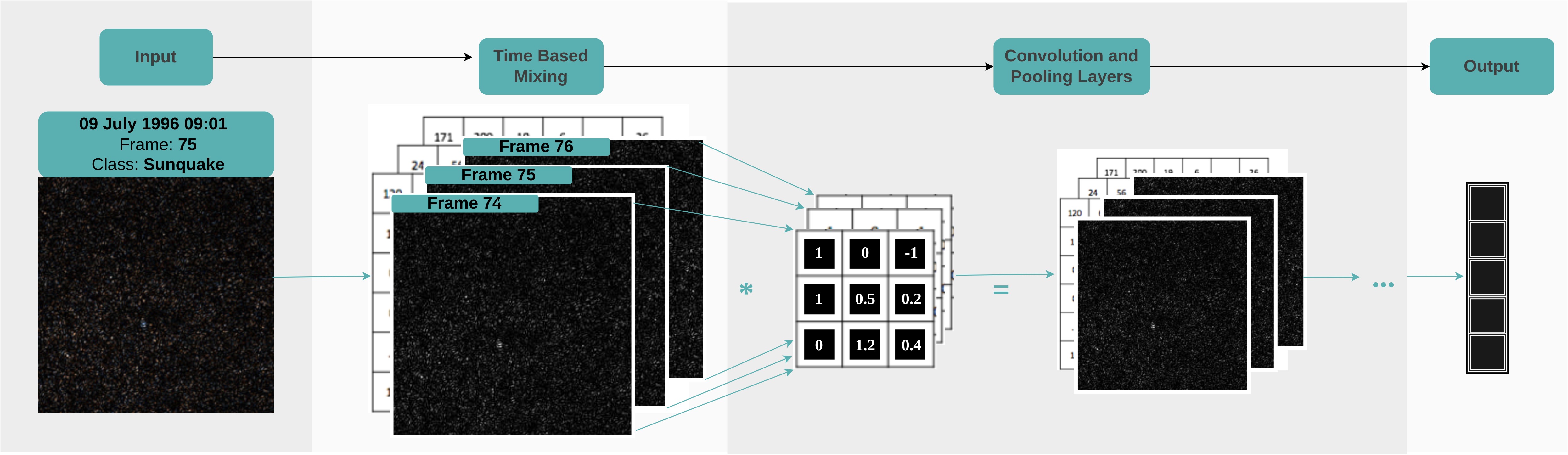}
\caption{Simplified view of how convolutional filters are applied to the input throughout a CNN based network. Starting leftmost, each input frame is rebuilt using the custom Time Based Mixing transform using a three window input. Then, Convolutional and pooling layers are applied to the transformed image. The rightmost output data product is flattened to the target latent dimension.}
\centering
\label{fig:cnn_prevs}
\end{figure}

For each model-input gray-scale sample $i$, a three dimensional image is created, where the first dimension, that typically represents the channels, stores the sample at index $i-1$, $i$ and $i+1$. In this way, the convolution and pooling filters applied to the sample  throughout the models' network capture the evolution of the sample over a time period of three steps. A visualization of this process is presented in Figure \ref{fig:cnn_prevs}. The models described in  Section \ref{s-results} make use of data constructed in this manner. This attempt of incorporating the time component significantly improves results (see Table \ref{table:macro_avf_high_prec_baseline}).

\subsection{Object Detection}

\label{sss-objectdet}
Because we currently do not employ additional methods for the explainability of the contrastive models described in Section \ref{sss-contrastive},  the location of predicted acoustic-emission sources and sunquakes is difficult to infer. Thus, an additional Faster Region-Based CNN (R-CNN) \citep{fasterrcnn} based Object Detection (OD) model is introduced to facilitate interpretation of the model outputs. 

The OD model predicts regions as box coordinates around sunquakes in single frame samples. We note that this model is not related or linked to the CL models described above, and its sole purpose is that of facilitating potential locations for sunquakes in frames where the CL models predict features.

As the OD model is trained only on region-annotated positive samples, it predicts sunquake regions for the majority of samples at runtime. The role of predicting the time of occurrence of sunquakes falls to the contrastive model. By utilizing the predictions yielded by the CL model, positive frames (or positions in time) are identified. Then, the OD model is used for extracting potential regions (or locations) for these predicted frames, to reduce the manual effort when testing the model on unlabeled data.

\newpage
\section{Results and Interpretation}\label{s-results}

This section presents our most notable results and findings. Following the experiments described in Section \ref{sss-contrastive}, we train two CL sunquake detection models with different particularities as described below. The outputs of these models are combined with OD model outputs to derive both temporal and location information of sunquakes for additional datasets.

Due to the large size of the models and the long training time, cross validation is not performed, on account of  computational requirements. A locked random state is used at run time for reproducibility. We note that it is possible for the results relying on this form of data preparation to suffer changes on alternative data shuffles.

\subsection{CL Models}

In this section we provide the architecture and parameters of two CL models and analyze their predictions. Significant implementation features are shared by both models.

For both CL approaches, the combined SC23 and SC24 data are split during load in an 80\,/\,20\,\% manner into training and validation sets.

We apply the custom Time Based Mixing transform to the entire dataset, followed by the Custom Low Pass and Solarize transforms with a threshold of 50 and with a 50\,\% probability. Then, we randomly apply one single positional transform out of horizontal or vertical flip, 90\textdegree, 180\textdegree, or 270\textdegree\   random rotations, with a probability of 50\,\%. Finally, we apply the Random Erase transform as described above, generating up to eight gray erasing rectangles of varying proportions.

Before moving on to describing the models, we must emphasize the problem of external bias induction. This occurs when, willingly or not, the model is presented with additional information from the exterior that facilitates training but may impact the reliability of results. From our experiments, we identified two biases that impose a level of risk.

Firstly, when loading the data, if the entire dataset is shuffled before performing the split, different samples associated with the same event may appear in both training and test data, inferring external information to the model regarding previously seen ARs. 
    
Secondly, when upsampling with geometric transformations, even though they are also applied at runtime to all samples, a transformation bias is induced to the model, making it more inclined to predict sunquakes for geometrically transformed samples. When performing inference, no geometric transformations should be applied to the input data so as not to affect the reliability of the predictions. 

The impact of such biases is not fully clear. To explain this, we prepared the model in Section \ref{ss-model1}, which presents none of the above biases, to offer a clean overview of baseline capabilities. Sections \ref{ss-model3} and \ref{ss-falsepos} will describe an in-depth prediction analysis resulting from an impacted model. The analysis is performed on the additional datasets, denoted with ``-" in Table \ref{table:used_sunquake_events}. 

\subsubsection{High Precision and Accuracy CL Model with No Known Biases}
\label{ss-model1}
To mitigate the shuffle bias described earlier, we  group the input data by event and only perform shuffling after dividing the groups between train and validation data. This assures that both split sets are self-contained in terms of included ARs. To better illustrate, if we assume that the Nth frame of an event is present in the training set, then so are all other frames belonging to that same event, and none are present in the validation set. 

We train a self-supervised ResNet-18 ImageNet-pre-trained CL model for 500 epochs on the SC23 and SC24 datasets (using positive upsampling), followed by a weighed multi-head supervised contrastive model for another 100 epochs (without upsampling). For the supervised model we used an encoding size of 512, a projection dimension of 128, and a temperature of 0.07. The first head includes a linear layer, a ReLu activation function, and an output encoding layer, while the second head is used for performing and monitoring classification during training and consists of a linear layer. 

We then take the resulting encodings for the training and test sets and pass them through an extensive list of classifiers. Building upon the results shown in Table \ref{table:macro_avf_high_prec_baseline}, we further analyze the embeddings with respect to the predictions provided by the polygonal kernel SVC. When loading data to this model, we separate events entirely between train and test data, and we infer the contrastive encoding on raw images only, avoiding both the shuffle and the transformation biases. 
To further tackle imbalance, besides encapsulating weights in the contrastive loss according to Equation \ref{eq2}, we use Smote Augmentation \citep{JMLR:v18:16-365} on the training data. We synthetically upsample positives with a sampling strategy of 0.2, and down sample negatives with a sampling strategy of 0.75. By this, first the positive samples are increased so that they measure 20\,\% of the total count of the negative samples, and the negative samples are reduced until the positive count is equal to 75\,\% of the negative count, leading to 2753 negative samples and 2065 positive samples used for training the classifiers.

\begin{figure}[!t]
\includegraphics[width=\textwidth]{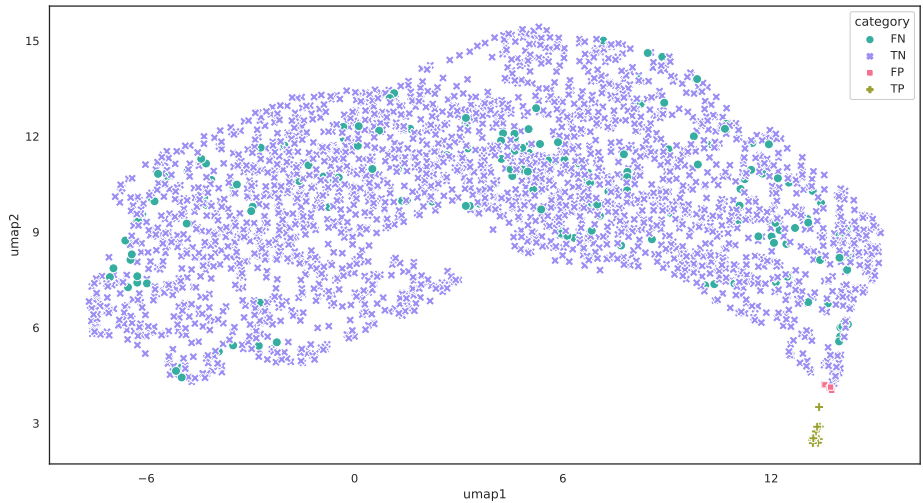}
\caption{The predictions of the High Precision and Accuracy model, on the test set (20\,\% of the SC23 and SC24 data), colored by their prediction correctness and clustered by UMAP components.}
\centering
\label{fig:high_precision_umap}
\end{figure}

Performance of this model before and after applying Smote Augmentation are provided in Table \ref{table:macro_avf_high_prec_baseline}. We see improvements of up to 20\,\% in precision,  and a 1\,\% accuracy increase in SVC (poly). Logistic Regression seems to not be affected by this augmentation, which may be explained by how the dense minority-class is distributed closely to the sparse majority-class, as depicted in Figure \ref{fig:high_precision_umap}.

\begin{table}[!t]
\tiny
 \setlength{\tabcolsep}{3.2pt}
\begin{tabular}{c c c c c c c c c c c }
 \hline
    
 Classifier &
    \multicolumn{2}{c}{K-NN (Bagging)} & 
        \multicolumn{2}{c}{SVC (Poly)} & 
            \multicolumn{2}{c}{SVC (RBF)} & 
                \multicolumn{2}{c}{Logistic Regression} & 
                    \multicolumn{2}{c}{SGD} \\ 
 \hline
Augmentation   & None & SMOTE & None & SMOTE & None & SMOTE & None & SMOTE & None & SMOTE \\
 \hline

Precision   & 0.63 & 0.65 & 0.66 & \textbf{0.84} & 0.49 & 0.59 & 0.64 & 0.64 & 0.54 & 0.62\\
Recall      & 0.54 & 0.54 & 0.54 & 0.54 & 0.50 & 0.54 & 0.54 & 0.54 & 0.54 & 0.54 \\
F1-Score    & 0.55 & 0.55 & 0.55 & 0.55 & 0.49 & 0.55 & 0.55 & 0.55 & 0.54 & 0.55 \\
Accuracy    & 0.93 & 0.93 & 0.93 & \textbf{0.94} & 0.93 & 0.92 & 0.93 & 0.93 & 0.89 & 0.93 \\ [1ex]
 \hline
 Metric Avg.    & 0.662 & 0.667 & 0.67 & \textbf{0.715} & 0.605 & 0.65 & 0.665 & 0.665 & 0.627 & 0.66 \\ [1ex]
 \hline
\end{tabular}
\caption{Macro Average performances of different classifiers over CL produced embeddings, trained with and without SMOTE augmentation, on the SC23 and SC24 test data (2622 negative and 186 positive samples)}
\label{table:macro_avf_high_prec_baseline}
\centering
\end{table}

Although Table \ref{table:macro_avf_high_prec_baseline} shows a high precision score, recall is small with only 20 predicted sunquakes samples, out of which 6 are false positives. The test set contains 11 sunquake events, listed in Table \ref{table:prediction_counts_baseline_precision}. As the table shows, seven events are recovered by this model.

\begin{table}[!t]
 \setlength{\tabcolsep}{5.1pt}
\tiny
        \begin{tabular}{c c c c c c c c c c  } 
             \hline
             Event & 
                \multicolumn{4}{c}{Counts} & Event &
                    \multicolumn{4}{c}{Counts}   \\ [0.5ex] 
     Date & TP & FP & FN & GT & Date & TP & FP & FN & GT \\
     \hline
    09 July 1996 09:01    & 0 & 0 & 19 & 19            &  04 July 2012 09:47	& 2	& 0	& 17 & 19 \\
    06 April 2001 19:13	& 0 & 0 & 16 & 16              &  06 July 2012 13:26	& 2	& 0	& 15 & 17 \\
    24 September 2001 09:35	& 0 & 0 & 11 & 11          &  08 November 2013 04:20	& 2	& 0	 & 18 & 20 \\
    23 July 2002 00:27    & 0	& 6 & 14 & 14          &  11 March 2015 16:11	& 2	& 0	& 14 & 16 \\
    05 March 2012 19:27	& 2	& 0 & 19 & 21              &  28 September 2015 14:53	& 2	& 0	& 18 & 20 \\
    06 March 2012 07:52	& 2	& 0	& 11 & 13              &  & & & \\
             \hline
        \end{tabular}
    \caption{TP, FP, FN sum for SMOTE augmented SVC (poly) predictions and positive GT count for each event in the SC23 and SC24 test data (2622 negative and 186 positive samples). In the case of events where $TP\equiv2$, the model predicts the exact transition frames into and out from the SQ}
    \label{table:prediction_counts_baseline_precision}
    \centering
\end{table}

To further analyze the particularities of the correct predictions and to identify the reason behind false positives, several characteristics of the embeddings are analyzed, beginning with the UMAP clustering, cosine and euclidean distances, and cosine similarities between consecutive frames. Using euclidean distance provides almost identical results. Findings indicate

that there is a considerable difference between the embeddings associated to data in SC23 and those of SC24, indicating either that models may recognize the different measuring instruments.

In Figure \ref{fig:high_precision_umap}, a UMAP plot is presented, depicting on the two axes, synthetically extracted features based on the contrastive embeddings.
The correctly predicted samples are visibly clustered at the tip of the other points, in the lower right-hand corner. FP are grouped together very close to TP, which is expected considering the use of a polygonal kernel during classification. 

At a first glance, an evident issue with the embedding clustering lies in the distribution of FN samples, which are randomly distributed alongside the TN. 

\begin{figure}[t]
\includegraphics[width=\textwidth]{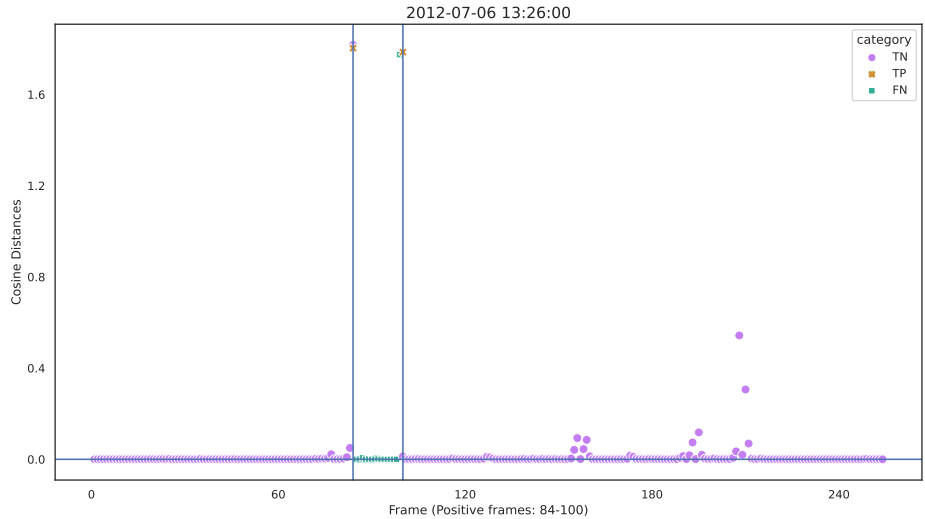}
\caption{Cosine Distances computed between consecutive frames' embedding for the 06 July 2012 13:26 event in the SC23 and SC24 test data, colored by prediction correctness. The $x$-axis is represented by the frame index for this event, and the $y$-axis denotes the Cosine Distance value.}
\label{fig:cosine_distances.png}
\centering
\end{figure}

Figure \ref{fig:cosine_distances.png} shows the measured Cosine Distances between consecutive frames, as a non-outlying plot of a successfully recovered sunquake in the 06 July 2012 13:26 event from the test set. We observe that the predicted frames corresponding to the recovered events are the leftmost and rightmost margins of the sunquake. A maximal Cosine Distance between the embedding vectors outputted by the CL component of our model appear between frames 83\,--\,84 and 100\,--\,101. The colors indicate the model's predictions with respect to each frame's embedding vector. These highly distanced frames are exactly the human-identified transitions into and out of a sunquake. This high cosine distance is considered to be the main reason behind the model's positive prediction for these transition frames. We note that all of the recovered sunquakes in SC24 maintain roughly the same characteristics.

We attempt to justify this behavior by pinpointing that for this experiment, when training the CL ML model, each input data sample is augmented with the custom Time-Based Mixing transform. We hypothesize that because of this, the embedding is capable of capturing a gradient in the intensity of the sunquake region between the channels of an individual sample. Further increasing the number of used channels might improve this result.

To understand why the immediately nearby frames located right before and after a sunquake are not also predicted as positives in spite of the visibly large cosine distance that they also present, we look into means and medians, sample level standard deviation, and embedding vectors difference from the mean. All these characteristics indicate a typical behavior for immediate pre- and post-quake frames as compared to other event frames, and a more evident discrepancy for quake marginal frames.

\begin{figure}[t]
\includegraphics[width=\textwidth]{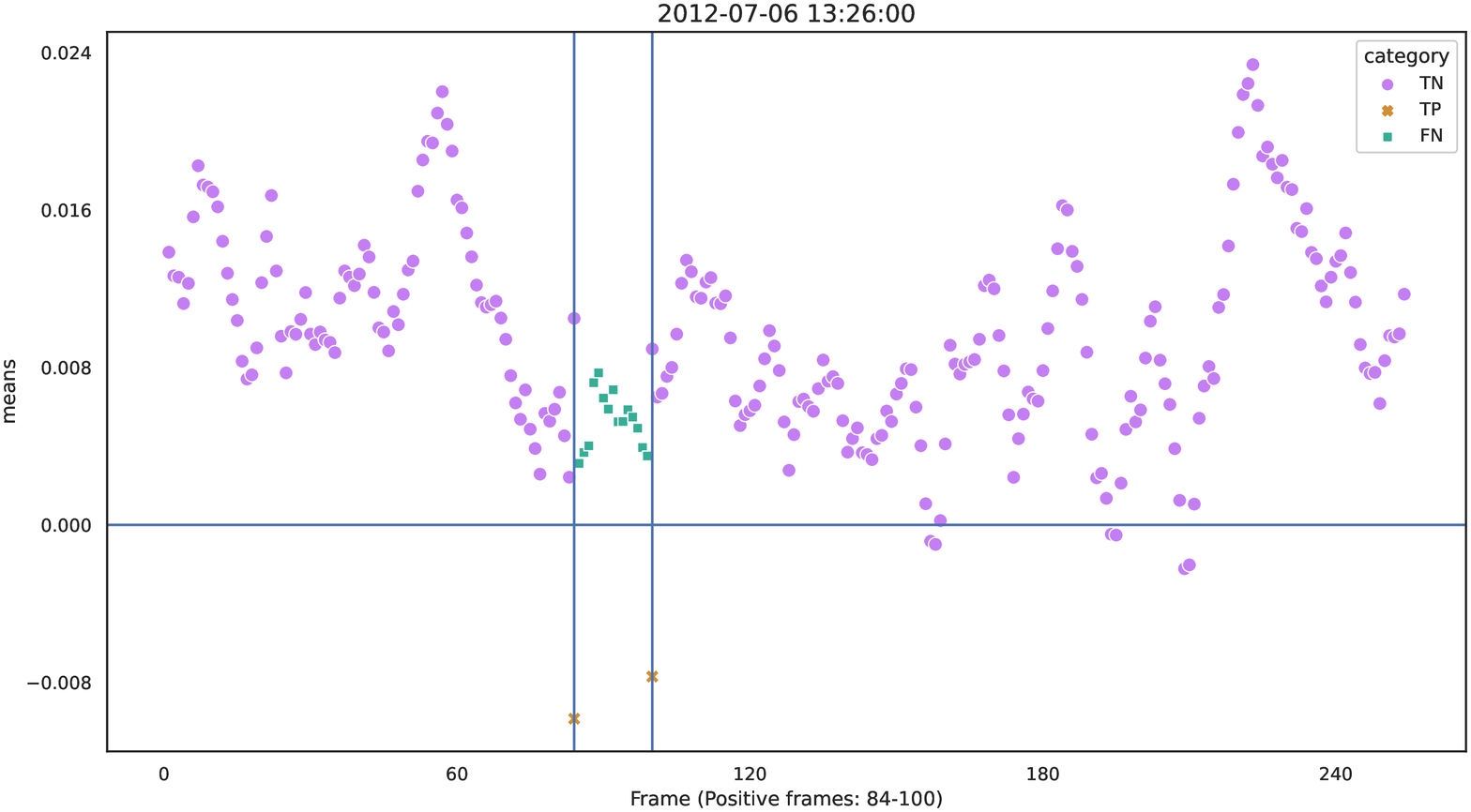}
\caption{Means for each frame embedding vector for the 06 July 2012 13:26 event in the SC23 and SC24 test data, colored by prediction correctness. The $x$-axis is represented by the frame index for this event, and the $y$-axis denotes the embedding mean value.}
\centering
\label{fig:means_embbeddings.png}
\end{figure}

An example of this behavior is depicted in Figure  \ref{fig:means_embbeddings.png}. 

We can easily see that the TP embeddings have a much lower mean than the adjacent ones. This behavior is also occurring in the other characteristic plots, but to a less evident extent.

There are four events for which the model is not able to identify correct sunquake frames, as seen in Table \ref{table:prediction_counts_baseline_precision}. However, the model does show common behavior when marking positives with respect to the cosine distances between consecutive frames. For example, in the 23 July 2002 00:27 event, the model predicts a shift in the gradient intensity at frames 170\,--\,173 and 233\,--\,235, but this is not due to any known sunquake. This can result either from an abnormal noise pattern, or from events that were not visually identified by human observers.

Lastly, we apply this model to the data associated to the additional datasets marked with ``-" in Table \ref{table:used_sunquake_events}. For these four unlabeled events, the predictions are as follows.

For 08 May 2012 13:02, one sunquake is predicted, around frames 180\,--\,188. This will be analyzed in more detail below. For 30 December 2011 03:03, a significant number of frames are predicted as sunquakes, which, given the high precision of this model, indicates that the dataset might be too different from those previously seen. Visually, the AR does not seem that clear with respect to the noisy quiet-sun area around it. One sub-interval of positive predictions is analyzed in depth below. For 10 May 2012 04:11 and 25 September 2011 08:46 no sunquakes are detected. Each of the additional sets should have observation of one Sunspot, as retrieved from our literature source \citep[][Table 1]{sharykin2020}.

\subsubsection{High Metrics CL Model with Shuffle Bias}
\label{ss-model3}

\begin{figure}[t]
\includegraphics[width=\textwidth]{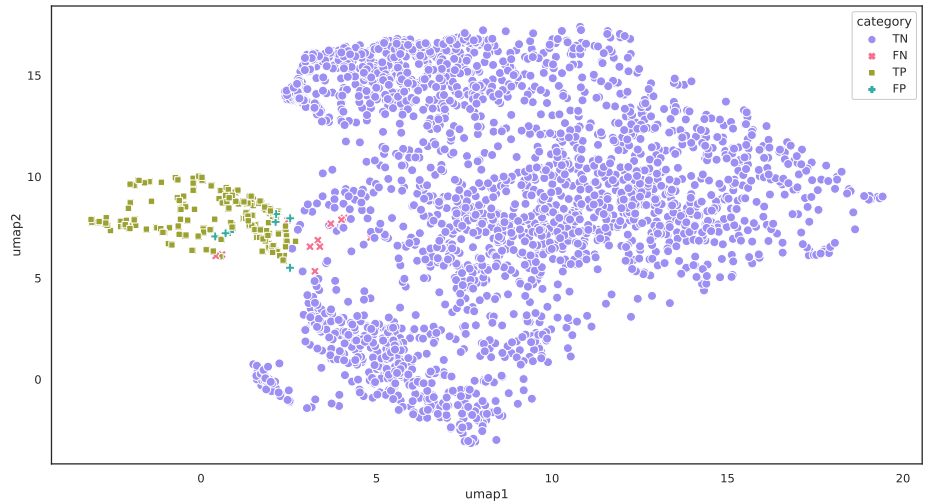}
\caption{ The predictions of the High Metrics with Shuffle Bias model, on the test set (20\,\% of the SC23 and SC24 data), colored by their correctness and clustered by UMAP components.}
\label{fig:umap_shuffled.png}
\centering
\end{figure}

We train a Supervised CL model with a DenseNet-121 ImageNet-pre-trained backbone, batch normalization, dropout, global average pooling, temperature of 0.1 and an encoding and projection of size 20 on the SC23 and SC24 dataset (using positive upsampling) for $\approx$50 epochs. This model encapsulates the shuffling bias, in that it is trained on the fully shuffled SC23 and SC24 dataset, at sample level. The encoding produced by this model is used to perform the classification. For this, similar to the previous model, an SVC classifier with a polygonal kernel is chosen.

This model comes as a significant improvement to its predecessor in terms of metrics, which are 
presented in Table \ref{table:macro_avf_shuffle_baseline}. Despite the boost, this model introduces a level of uncertainty due to the present bias related to data shuffling. To train the model, we modify the data loader so that data are no longer grouped by their corresponding event date before the split. Hence, after shuffling, the training and validation set may contain frames belonging to the same initial cube, such as the Nth frame belonging to event 06 April 2001 19:13 residing in the training set, and the $N+K$th frame residing in the test set. This modification facilitates the model's ability to produce meaningful embeddings for test data, as the same AR might be present in the training set samples. 

Figure \ref{fig:umap_shuffled.png} shows clustering of the embeddings produced by the model for the test set. The distribution is quite sparse, but there is a clear separation between the positive and negative class. Falsely predicted samples are tightly coupled in between both clusters, thus making it difficult for the model to clearly classify them.

\begin{figure}[t]
\includegraphics[width=\textwidth]{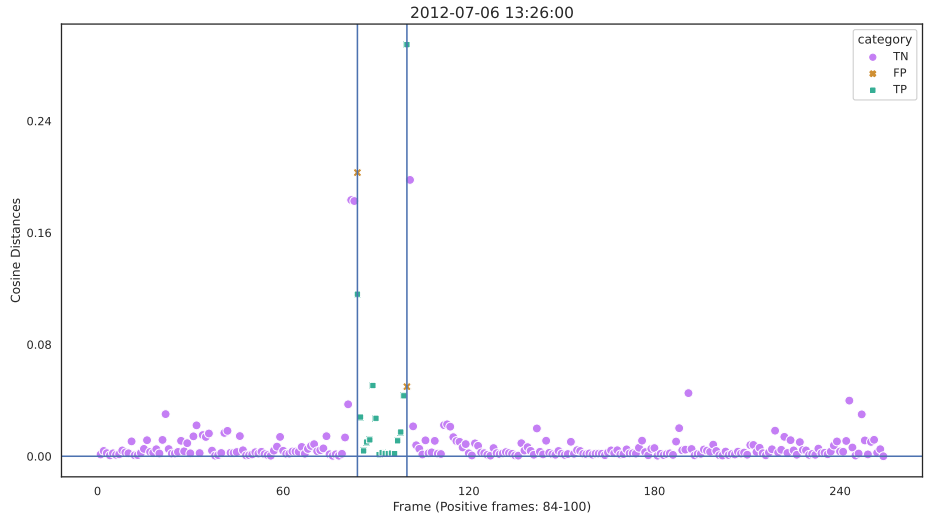}
\caption{Cosine Distances computed between consecutive frames' embeddings vectors for the 06 July 2012 13:26 event in the SC23 and SC24 test data, colored by prediction correctness.}
\label{fig:cosine_distances_shuffled.png}
\centering
\end{figure}

In spite of the shuffle bias this model presents plots of cosine distances between consecutive frame embeddings, as that in Figure \ref{fig:cosine_distances_shuffled.png}, maintain the characteristics of those in the high-precision model described earlier in Section \ref{ss-model1}. 

Common behavior displayed between the two different models indicates that aspects such as embedding means and cosine distances are relevant during the process of learning the given task. It also validates the assumption that by using the custom Time-Based Mixing augmentation described in Section \ref{sss-augmentations}, transitions to and from a sunquake are adequately captured by CL models.

This particular model presents no weighting at the CL level, indicating that models are much more easily trained on sample-level shuffled data. Even though this model appears to lack imbalance impediments, we provide a comparison between classification with and without SMOTE augmentation in Table \ref{table:macro_avf_shuffle_baseline}. An improvement in F1-Score up to 30\,\% is noted for K-NN and SVC (RBF), but Logistic Regression seems to suffer from the augmentation, with a 6\,\% decrease in precision, as analogous to the no performance gains observed for logistic regressors in Section \ref{ss-model1}. The SVC (poly) which is the top average scoring SMOTE-based classifier gains a small 1\,\% improvement in precision and recall.  

\begin{table}[t]
\tiny
 \setlength{\tabcolsep}{3.15pt}
\begin{tabular}{c c c c c c c c c c c }
 \hline
    
 Classifier &
    \multicolumn{2}{c}{K-NN (Bagging)} & 
        \multicolumn{2}{c}{SVC (Poly)} & 
            \multicolumn{2}{c}{SVC (RBF)} & 
                \multicolumn{2}{c}{Logistic Regression} & 
                    \multicolumn{2}{c}{SGD} \\ 
 \hline
Augmentation   & None & SMOTE & None & SMOTE & None & SMOTE & None & SMOTE & None & SMOTE \\
 \hline

Precision   & 0.97 & 0.80 & 0.97 & 0.98 & \textbf{0.99} & 0.94 & 0.97 & 0.91 & 0.91 & 0.89\\
Recall      & 0.54 & 0.96 & 0.96 & 0.97 & 0.86 & \textbf{0.99} & 0.98 & \textbf{0.99} & 0.94 & 0.93 \\
F1-Score    & 0.55 & 0.86 & 0.97 & 0.97 & 0.91 & 0.96 & \textbf{0.98} & 0.95 & 0.94 & 0.93 \\
Accuracy    & 0.94 & 0.95 & \textbf{0.99} & \textbf{0.99} & 0.98 & \textbf{0.99} & \textbf{0.99} & \textbf{0.99} & 0.98 & 0.98 \\ [1ex]
\hline
Metrics Avg.    & 0.750 & 0.892 & 0.947 & \textbf{0.977} & 0.935 & 0.97 & \textbf{0.98} & 0.96 & 0.942 & 0.932 \\ [1ex]
 \hline
\end{tabular}
\caption{Macro Average performances of different classifiers over CL produced embeddings, trained with and without SMOTE augmentation, on the SC23 and SC24 test data (2622 negative and 186 positive samples)}
\label{table:macro_avf_shuffle_baseline}
\centering
\end{table}

\subsection{OD Model}

This section provides the validation results of the OD model for both SC23 and SC24 test data. The model was trained on 80\,\% of the positive events in SC23 using a Faster R-CNN architecture and a ResNet-50 backbone, pre-trained on ImageNet. This selection is due to the limitations of MDI. The instrument was sensible enough to only capture stronger sunquake events, leading to a selection bias in which most detected events have good signal-to-noise ratio in the egression power maps. For this localization task, this choice has the advantage producing a clean qualitative dataset, but also the disadvantage of incompleteness. 

The most commonly used validation metric in OD is the Intersection over Union (IoU), which quantifies the degree of overlap between the predicted region box and the GT. Table \ref{table:od_performances} shows four types of metrics. The IoU metric results (i) on our data appear underwhelming. This is because the manually annotated regions vary in size between events, oftentimes including padding. For this reason, we analyse the detections using additional metrics that better capture the desired outcome of this model. We look at: correct signature coverage (ii), the overlap of averages over different minimum sunquake duration (iii), and the percentages of predicted boxes inside the GT (iv). Additional information and visualizations on decreased IoU values in the case of correct detections for the test data in SC24 are provided in Appendix \ref{od-appendix}. 

For SC24, the predicted IoU and the GT boxes overlap by $\approx$21.5\,\%. An average coverage of correctly localized sunquake signatures in 44.2\,\% of the total positive frames for singular events. This means that although sunquake locations are recovered, not all consecutive frames corresponding to one event are successful in capturing the signal. By manually reviewing predictions, we found that the model tends to perform better for non-marginal sunquake frames, supported by the fact that a significant part of the training data contains stronger examples.  Moreover, we test introducing the required minimum duration of predicting the sunquake signature in the same spot. We find that  while increasing the minimum duration time, the average percentage of identified signatures for singular events decreases down to 39.3\,\%. 

We assume the 44\,\% and 62.6\,\% event-level average of correctly marked signature regions for SC24 and SC23, respectively, to be sufficient for our current goal of enhancing the CL model predictions with a probable location component. Importantly, we note that the currently described OD model should not be considered adequate as a standalone detection tool.

\begin{table}[t]
\tiny
\begin{tabular}{c c c c }
 \hline
    
 & Metric & SC23 Data & SC24 Data \\
 \hline
 
 (i)   & IoU avg. & 29.5 \,\% & 21.5\,\%  \\
 (ii)  & correct-signature coverage  & 62.6\,\% & 44.2\,\% \\
 (iii) & min. duration $1$ frames: correct-signature detections avg. & 100\,\% & 84.8\,\% \\
 (iii) & min. duration $4$ frames: correct-signature detections avg. & 75\,\% & 63.6\,\% \\
 (iii) & min. duration $6$ frames: correct-signature detections avg.& 75\,\% & 45.4 \,\% \\
 (iii) & min. duration $8$ frames: correct-signature detections avg.& 75\,\% & 39.3\,\% \\
 (iv)  & frame-level predicted boxes included in GT & 30\,\% & 20\,\% \\  
 
 \hline

\end{tabular}
\caption{Event-level OD Model performances for capturing sunquake signature boxes for the positive samples in SC23 and SC24 test sets.}
\label{table:od_performances}
\centering
\end{table}

\newpage
\subsection{An Analysis of Sunquake Detections on Additional Datasets}
\label{ss-falsepos}

\begin{figure}[ht]
\includegraphics[width=\textwidth]{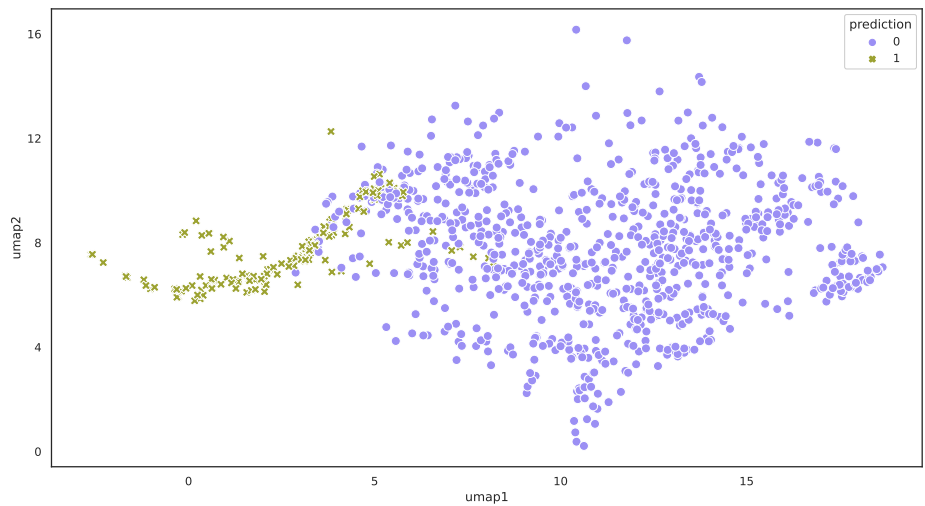}
\caption{The predictions of the High Metrics with Shuffle Bias model, on SC24 additional dataset, colored by their prediction value and clustered by UMAP components. 0 represents a negative prediction and 1 represents a sunquake prediction. }
\label{fig:umap_uncertain.png}
\centering
\end{figure}

For the aforementioned ``-" events in Table \ref{table:used_sunquake_events}, the CL model predicts a total of 93 positive frames and 923 negatives. The OD model described in Section \ref{sss-objectdet} is then applied to the positively predicted frames to extract potential regions. Figure \ref{fig:umap_uncertain.png} shows the clustering of embeddings, colored by prediction.  Although slightly different in distribution, the UMAP is quite consistent in interpretation to the test set clustering shown in Figures \ref{fig:high_precision_umap} and \ref{fig:umap_shuffled.png} discussed above.

\begin{figure}[!t]
\includegraphics[width=\textwidth]{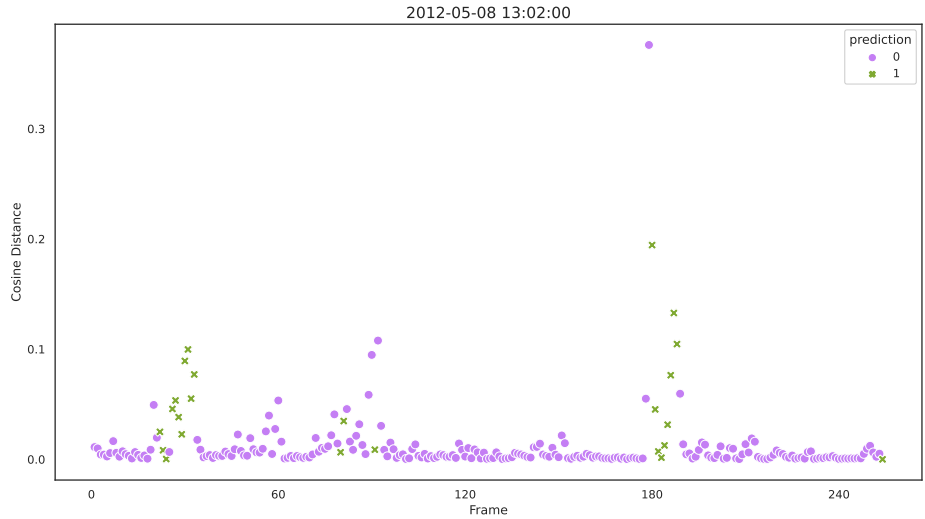}
\caption{Cosine Distances computed between consecutive frames' embedding vectors for the 08 May 2012 13:02 dataset of SC24, colored by their prediction value. 0 represents a negative prediction and 1 represents a sunquake prediction. }
\label{fig:cosine_uncertain2012.png}
\centering
\end{figure}

The predictions are as follows:
For the two datasets of 10 May 2012 04:11 and 25 September 2011 08:46 frame 199 is marked, and frames 87\,--\,89 are marked as sunquakes, respectively. Per our identification and selection criteria, one and respectively three frames are insufficient to justify a sunquake signature. An analysis of the higher-atmosphere data showed no candidate  eruptions.
With respect to the 08 May 2012 13:02 dataset, two sunquakes are predicted at frames $[22\,-\,33]$ and $[180\,-\,180]$ in different locations. A cosine-distances plot is shown for this event's embeddings in Figure  \ref{fig:cosine_uncertain2012.png}. As similar to the test events, the medium-high and high values of this characteristic for sunquake margins are maintained for both detections, respectively. The first prediction proved to be a false positive. The second sunquake is predicted by both CL models, and it is further scrutinized below.
The 30 December 2011 03:03 dataset contains six sunquake identifications. Five cases either proved inconsistent with the temporal detection criteria, or the OD did not successively converge on the same location for the entire CL duration. The last detected event $[15\,-\,22]$ is given further consideration below.

We aimed to enhance the level of explainability of the contrastive model, in the absence of other implemented methods, by additionally utilizing our most impactful augmentation, the Solarized Low Pass Filter custom transform, alongside the OD approach. 

\subsubsection{08 May 2012 13:02 Prediction Analysis}
\begin{figure}[t!]
\includegraphics[width=\textwidth]{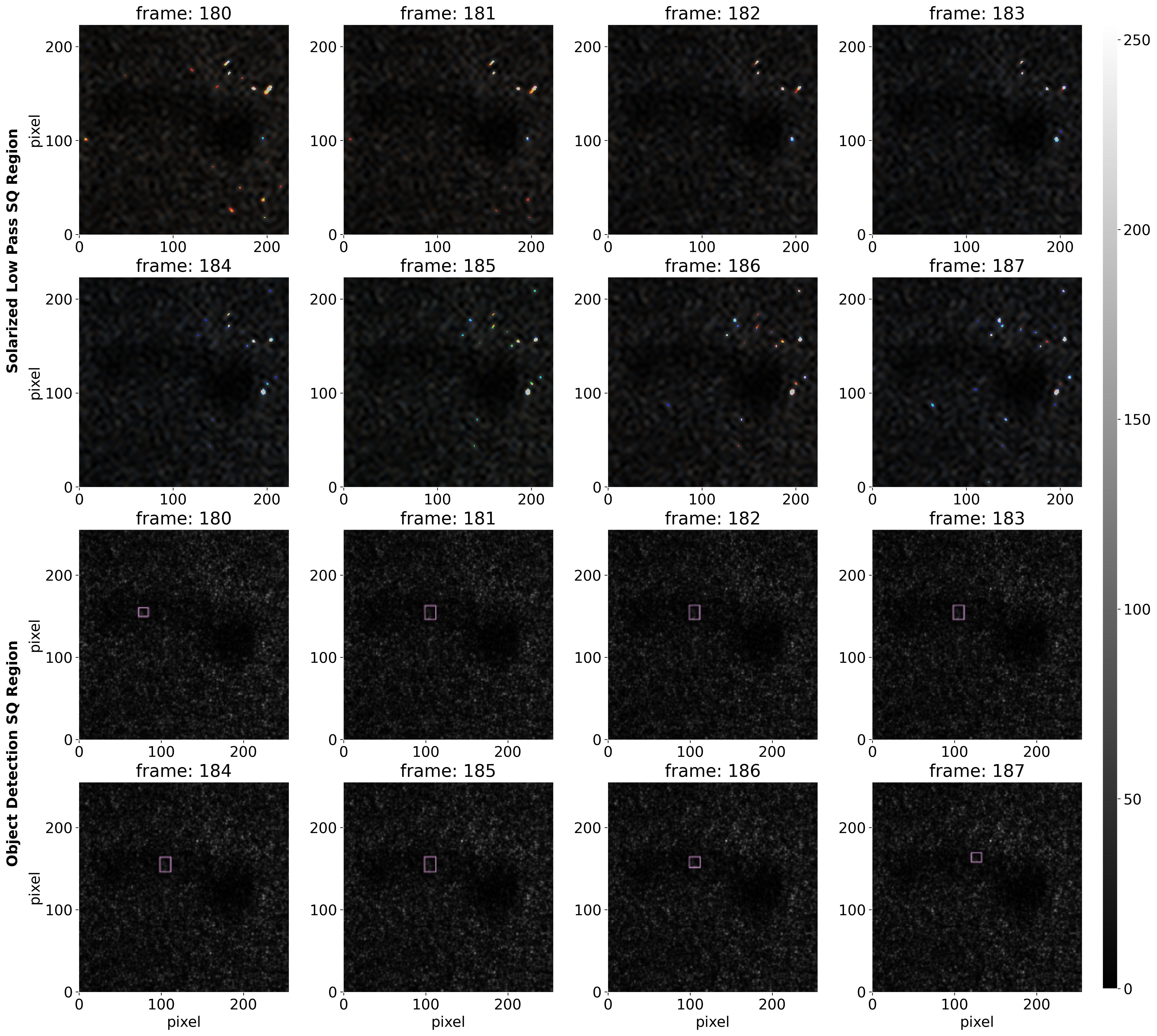}
\caption{Position of identified acoustic signatures based on Solarized Low Pass Filter and OD Regions for dataset 08 May 2012 13:02, at positively predicted frames [180\,--\,188). 8-bit intensity scaling is used.}
\centering
\label{fig:detections_and_lowpass_2012_05_08_event}
\end{figure}

In Figure \ref{fig:detections_and_lowpass_2012_05_08_event} we observe the predicted 08 May 2012 13:02 sunquake at frames $[180\,-\,188]$ using both detection approaches. Both models described in Sections \ref{ss-model1} and \ref{ss-model3} predict a sunquake occurrence in this temporal interval. Starting from this, a manual review of the respective regions is performed to identify sunquake presence and evaluate the behavior of the algorithms used. The OD identification shows a feature at position $[80\,,\,135]$ (Figure \ref{fig:detections_and_lowpass_2012_05_08_event} rows 3 and 4). The detection marked with a purple box maintains a fixed position starting from Frame 181, close to the center of the AR complex. The Solarized Low Pass Filter presents a spot of gradually increasing intensity on the right-side region at position $\approx\,[200\,,\,100]$ (Figure  \ref{fig:detections_and_lowpass_2012_05_08_event} rows 1 and 2). The other features are too short-lived to be classified as sunquakes. These are at two completely different locations, where each appears consistent temporally with the CL model detection. 

To disentangle this aspect, we use higher atmosphere observations from SDO's \textit{Atmospheric Imaging Assembly} \citep[AIA:][]{lemen2012} and from the \textit{Reuven Ramaty High-Energy Solar Spectroscopic Imager} \citep[RHESSI:][]{lin2002} to probe the eruptive and high-energy signatures that are usually associated to sunquake activity. The AIA data calibrated to level-1 data are obtained from the JSOC (\href{http://jsoc.stanford.edu/ajax/exportdata.html}{\textsf{jsoc.stanford.edu}}) around the predicted acoustic-source time intervals. The RHESSI high-energy source location is computed using the Clean algorithm \citep{Hurford+2002} by integrating the signal in the 6\,--\,12\,eV range measured around times of the maximum X-ray emission for each event.

\begin{figure}[ht]
\includegraphics[width=\textwidth]{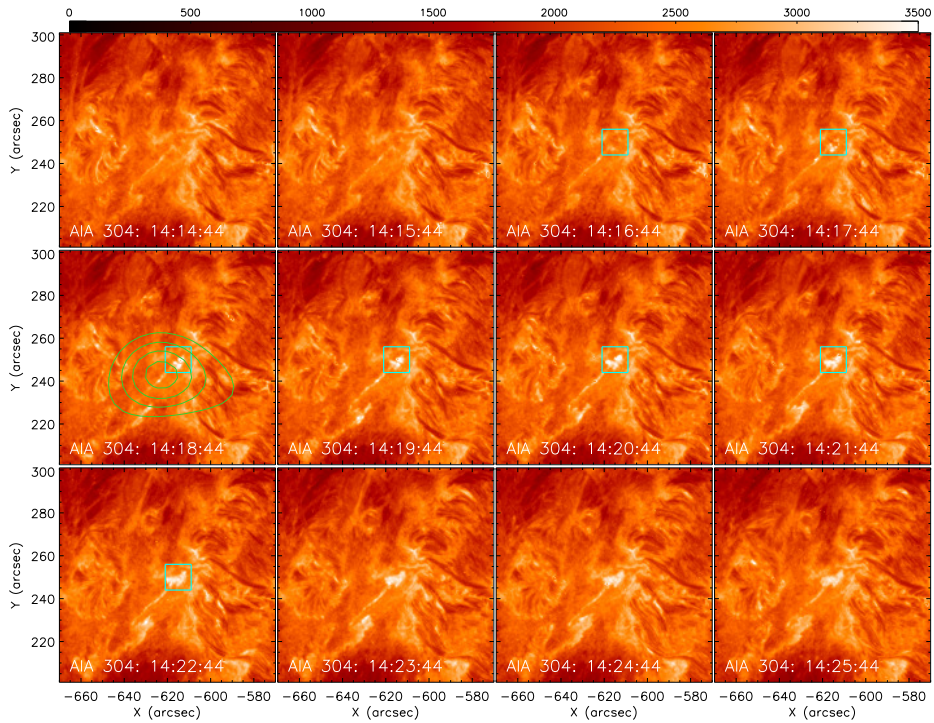}
\caption{Flaring activity as seen in the AIA 304\,\AA\ channel related to the 08 May 2012 13:02 dataset. The observation time for each AIA frame is included at the bottom of each frame. For the temporal interval when the OD kernel is identified, its position is marked by a \textit{cyan box} in the corresponding frames. The RHESSI high-energy X-ray 6\,--\,12\,eV signature location is shown as the \textit{green contours} for the time of the maximum flaring.}
\centering
\label{fig:20120508_map}
\end{figure}

\begin{figure}[ht]
\includegraphics[width=\textwidth]{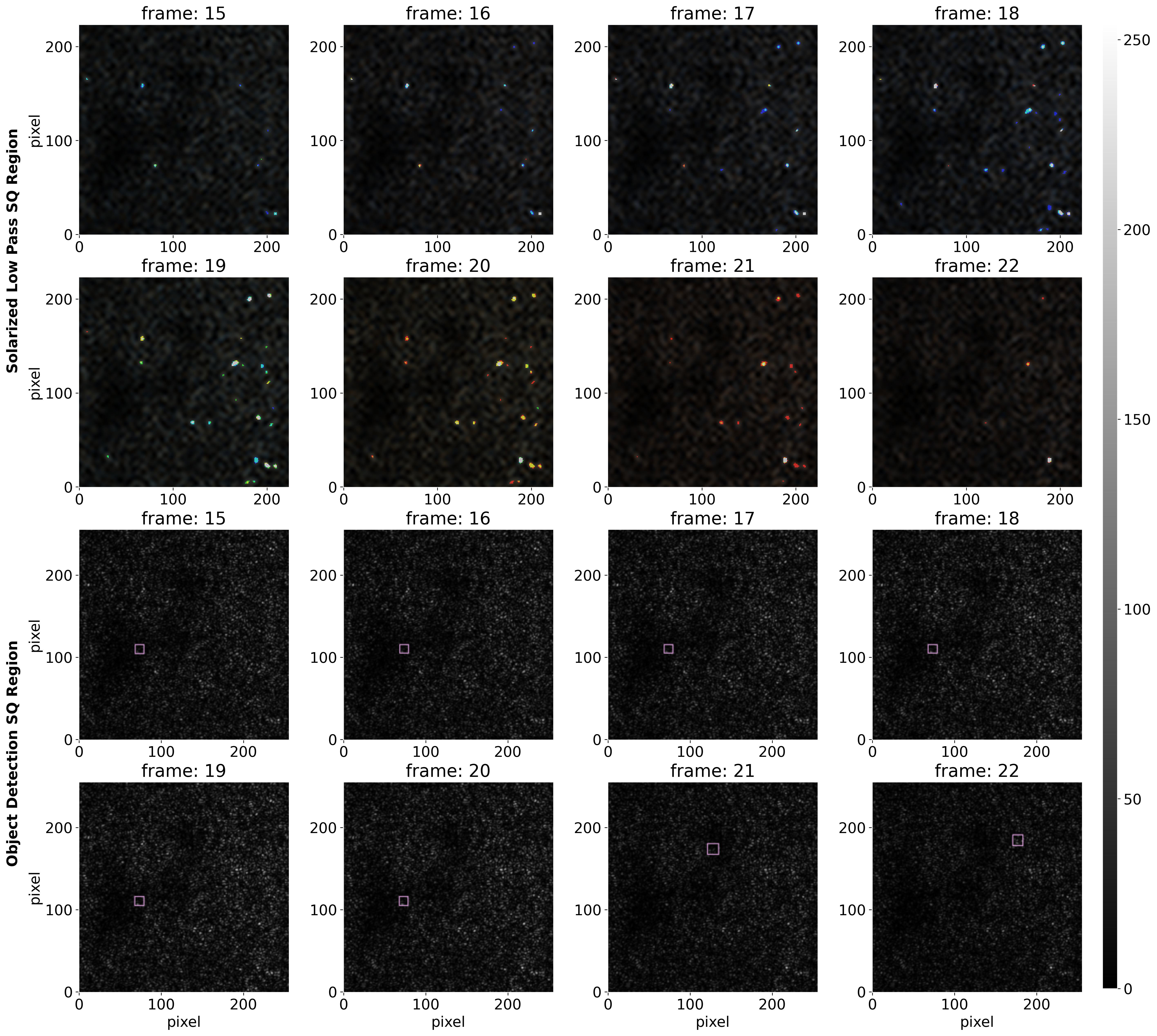}
\caption{Position of candidate sunquake signatures based on Solarized Low Pass Filter and OD Regions for dataset 30 December 2011 03:03, at positively predicted frames [15\,,\,23]. 8-bit intensity scaling is used.}
\centering
\label{fig:detections_and_lowpass_2011_12_30_event}
\end{figure}

We explore the emission of the solar atmosphere during the times indicated by the OD kernel detection. For this flaring event, the most significant signatures are observed in the AIA 304\,\AA\ emission originating in the solar chromosphere and transition region, as presented in Figure \ref{fig:20120508_map}. The OD detected position (cyan square) appears to overlap the footpoints of a mostly chromospheric flaring event.
The weak RHESSI X-ray 6\,--\,12\,eV source (green contours) is found to match the location of the AIA flaring. We note that the flaring appears to be visible for more frames, beyond the OD-marked interval. This is consistent with the fact that the signal in the egression power map is generated by the impulsive initiation of the flaring, while afterwards the chromosphere continues to radiate the generated energy. The second location inferred from the Solarized Low Pass Filter is discarded, as we could not find any clear eruptive or high-energy manifestation that can be associated with this location and time. 

\subsubsection{30 December 2011 03:03 Prediction Analysis}

\begin{figure}[ht]
\includegraphics[width=\textwidth]{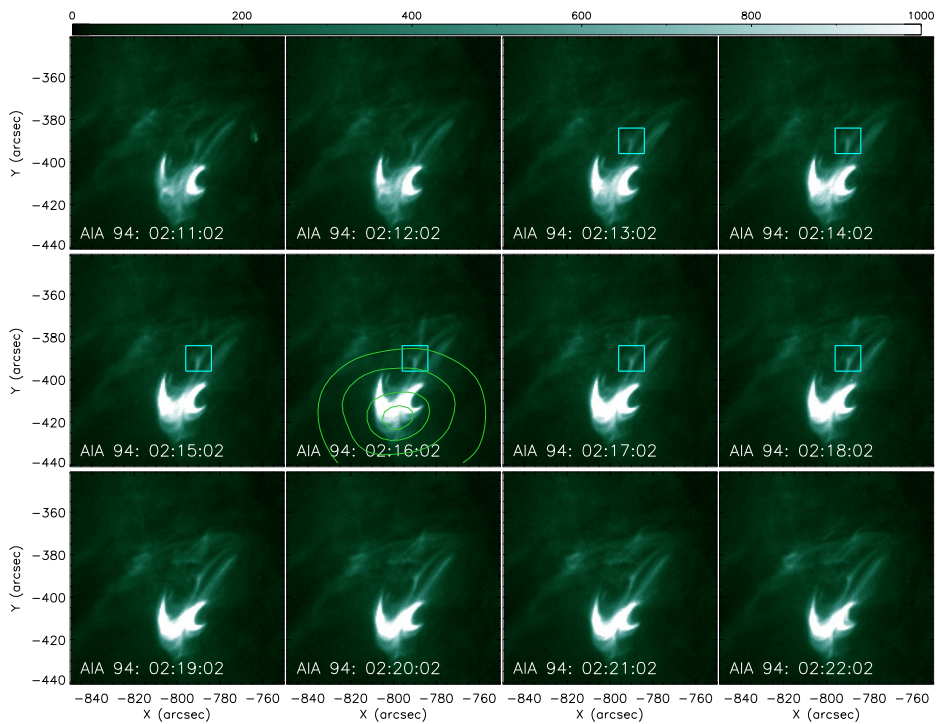}
\caption{Flaring coronal loops as seen in AIA 94\,\AA\ channel related to the 30 December 2011 event, with the position of identified OD kernel marked by the \textit{cyan box} and the RHESSI high-energy 6\,--\,12\,eV signature location shown as the \textit{green contours}.}
\centering
\label{fig:20111230_map}
\end{figure}

For our second example dataset, 30 December 2011 03:03, the Solarized Low Pass Filter is shown in Figure \ref{fig:detections_and_lowpass_2011_12_30_event}, while Figure \ref{fig:20111230_map} shows the hot AIA 94\,\AA\ channel in which most of the emission is recorded for this particular flare. The locations of the OD kernel (cyan) and the RHESSI source (green) are also included. This event is occurring very close to the solar limb, so projection effects are non-negligible in both the AIA emission and in the data used for acoustic-signature identification. The egression-power maps are de-projected to remove the solar rotation, while the AIA and RHESSI are significantly influenced by projection effects. In addition, the egression-power maps are mapping the solar photosphere, while the AIA 94\,\AA\ channel is mapping the very high corona. Thus, we explain the small mismatch of about $10^{\prime\prime}$\,--$\,15^{\prime\prime}$ between the source in the egression-power maps and the AIA and RHESSI data as a product of superposing all these effects. 
In this case, the Solarized Low Pass Filter shown in Figure \ref{fig:detections_and_lowpass_2011_12_30_event} did not capture our small acoustic region of interest, and has not identified other stronger acoustic sources with sufficient lifetimes for consideration.
We note that the insufficient frames where the OD has consistently detected the flaring location ($\approx$300 seconds), makes this event not fully compatible with a sunquake identification. This aspect, when coupled with the small location mismatch and the high projection effects of the observation, makes this association less strong than in the case of the 08 May 2012 13:02 dataset, but still relevant, at least with respect to qualitative and prospective application criteria.

\subsubsection{Discussion}

In both examples, we observe the AIA and RHESSI data to show eruptions accompanied by high-energy X-ray emission with class $\approx \textrm{C}1.0$ at the approximate location of the OD source during the same temporal intervals of the CL predicted sunquake intervals in both CL models. We tentatively hypothesize, by conjecture, that these source locations might be desiderated weak acoustic emission signatures that are produced by less powerful eruptions, even weaker than the source discussed by \citet{Sharykin+2015}. 

\begin{figure}[t]
\centering
\includegraphics[width=\linewidth]{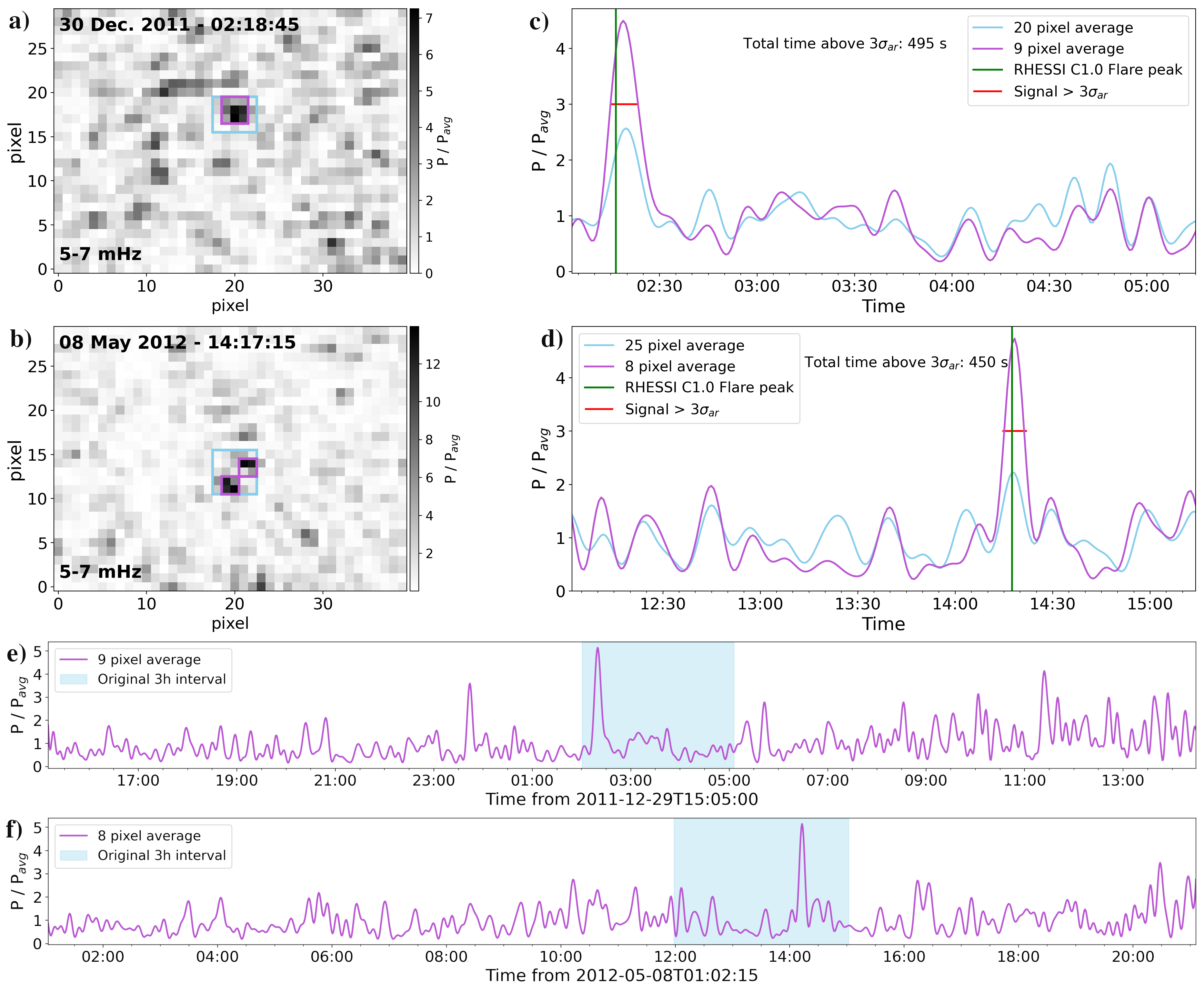}
\caption{\textbf{Panels a) and b):} Background normalized acoustic-emission power maps in the 4\,--\,7\,mHz band at the maximum of emission for the two detected weak sources. \textbf{Panels c) and d):} Evolution of the measured power with respect to background over two highlighted regions over our standard 3 hour dataset interval, one strictly including the acoustic kernels (\textit{purple}), the other also containing some background contribution (\textit{light blue}). The \textit{green} vertical line corresponds to the time of maximum RHESSI emission, and the horizontal \textit{red} line marks the interval when the retrieved signal is over the 3$\sigma$ level. \textbf{Panels e) and f):} The highlighted purple regions are tracked for longer temporal intervals to better constrain the background signal. The original 3 hour intervals are highlighted in blue.}
\label{fig:new_detections.png}
\end{figure}

Figure \ref{fig:new_detections.png} presents a more detailed analysis of the acoustic emission accompanying the AIA and RHESSI flares. The a) and b) panels show that both events are visually identifiable in egression power maps. The total emitted power over the three-hour background in ARs (P/P$_{avg}$) exceeded 7 and 14 in individual locations for the 31 December 2011 and 08 May 2012 events, respectively. The c) and d) panels show the P/P$_{avg}$ integrated over different kernel sizes over the three hours of observation of each AR. The more compact kernels (purple) maintain detection levels of above 4$\sigma_{ar}$ in both events discussed above. We note that detection limits would decrease even further to a $\approx 3\sigma_{qs}$ level if evaluating the temporal median signal in regions outside the less noisy AR shadows. 

Following the discussion of \citet{chen2021}, we assess that these detections represent very weak sources. In addition, we note that in general a modest $4\sigma$-like detection does not completely exclude us from the risk of spurious signal. To further constrain the detection, we additionally track the kernels over longer temporal intervals of the order of one day for each event (panels e) and f)).
Both discussed detections remain the most dominant features in the tracked regions, with detection levels reaching a more optimistic 5$\sigma_{ar}$ in this extended temporal noise statistics.  These concerns are also somewhat alleviated by the temporal consistency, and the sets of clustered pixels manifesting similarly. Lastly, when following the temporal evolution curves in panels c) and d) during the peak acoustic emission, we find that the 2011 event maintains a consistent $>3\,\sigma_{ar}$ detection for $\approx$\,495 consecutive seconds, on the edge of the detection threshold set by Equation \ref{eqdt}. The 2012 event falls marginally under the established threshold, where a $>3\sigma_{ar}$ detection is maintained for $\approx$\,450 seconds. Thus, although a set of stacking correlations between reasonable acoustic power, co-temporal AIA flaring, along with qualitatively overlapping RHESSI sources present a consistent set of evidence, we stress that these two aforementioned events would be hard to classify as bona-fide sunquakes when following a traditional holography based helioseismology analysis, as a number of methodological and statistical criteria (detection limit, temporal and spatial correlations, kernel size, temporal length, etc.) can hold up to only a qualitative level. These limitations can originate from any combination of instrumental, ML model, interpretation, or statistics effects.

We also remind the reader that our findings represent only a footing, upon which we encourage further exploration, as it is based on only two out of four analyzed datasets that contained one apparently positive example each. Preparing and ingesting a large dataset of acoustic emission of ARs that do not formally contain known sunquakes is required alongside more thorough detection limit statistics in order to confirm this result. Such a dataset should not be included in the training/validation and used only for highlighting temporal and spatial location of potentially unknown acoustic events. When using only HMI, such type of study can also help with establishing a lower flare energy detection limit when concerning ML applicability.

Although the Solarized Low Pass Filter improves detection metrics as has been demonstrated in Section \ref{sss-augmentations} and exemplified in Figure \ref{fig:solarized_low_pass_filter}, we find it is prone to introducing artifact regions into outputs when used in a stand-alone manner for explainability, as it is agnostic of our imposed criteria. For example, due to current holography limitations, the acoustic sources usually need to be inside the ARs to be identifiable. We thus deduce that although using this custom augmentation is tremendously beneficial to the overall training and validation of solar acoustic-emission datasets, careful consideration should be put into its explainability applications and sources that it might locate, as they have a significant chance to be FP. As a corollary, we note that this augmentation will increase the FP metrics of a trained model. This aspect again illustrates the complexity in interpreting these observations in a criteria agnostic ML fashion.

False detections such as the second location at frames $[180\,$--$\,188]$ in event 08 May 2012 13:02 indicate that our current models are not very robust to spurious correlations; for instance strong shifts in the gradient intensity in the ARs shadow, which are essentially noise, are predicted as sunquakes. To alleviate this, techniques that instruct the model to distinguish between false and correct sunquake signatures need to be employed. As a first step towards this, we plan to label our future datasets into multiple classes such as: strong SQ signature, weak SQ signature, AR shadow intensity shift (not SQ), etc. These labels can then be adopted in the CL methods to encourage producing sunquake representations that are dissimilar to their unfounded lookalikes.

We reiterate that both identified acoustic-emission sources, although promising as potential applications, do not present the actual position of sunquake identifications from the perspective of the CL models,
which only provide the temporal component. 
In the future, we will add the significant effort required to augment these trained models with a suite of CL-specific explainability features, which will be able to extract positions directly from the model output.

\section{Conclusion} 
      \label{S-Conclusion} 

\begin{figure}[ht]
\includegraphics[width=\textwidth]{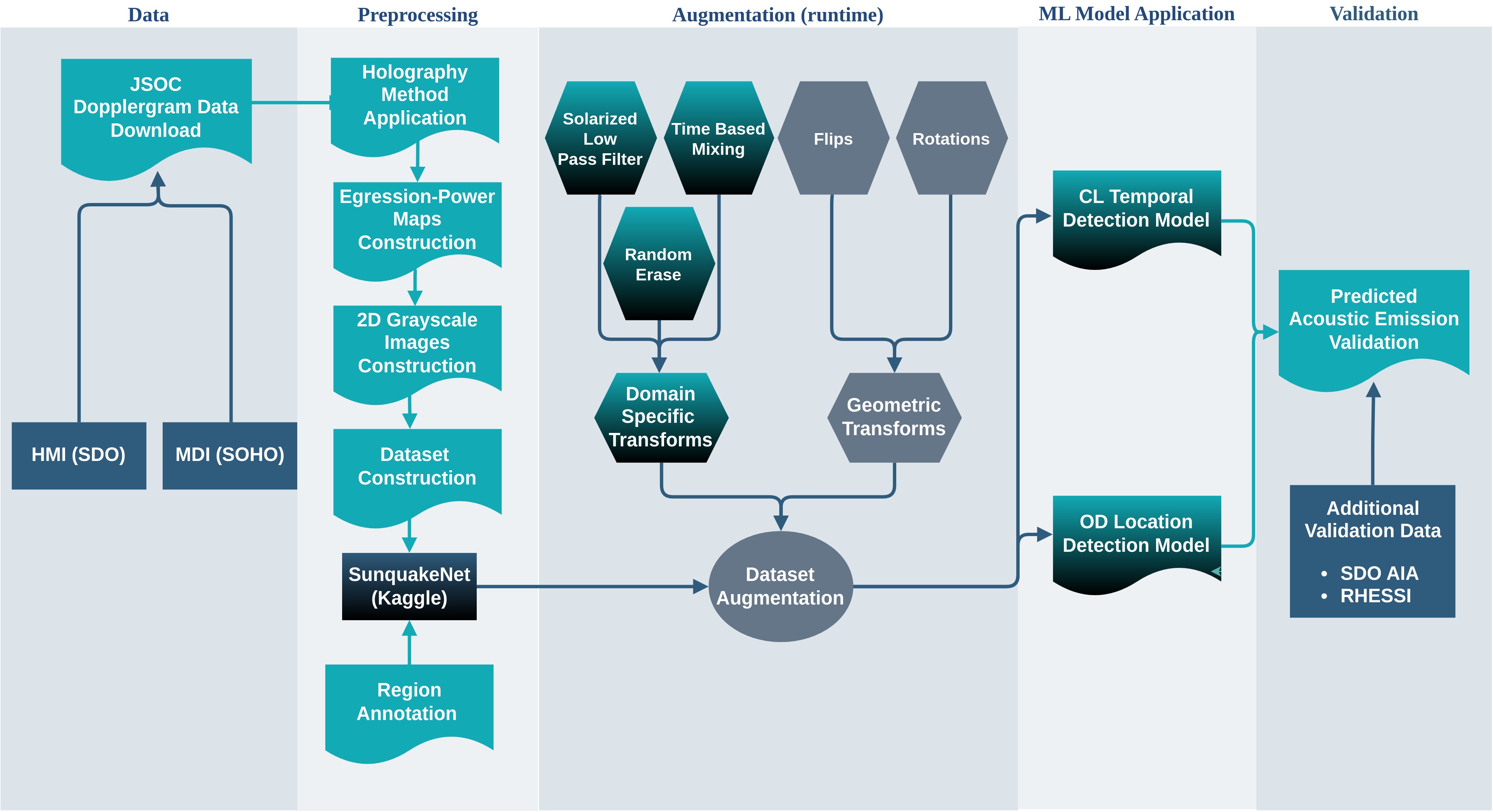}
\caption{Process diagram of proposed solution. The \textit{gray-labeled} augmentations represent standard ML approaches, while the others depict custom transforms that were found to improve the sunquake prediction model.}
\centering
\label{fig:process.png}
\end{figure}

In this work, we presented a pedagogical approach on the application of ML methodologies to acoustic emission image data of solar ARs. We constructed a curated dataset of major sunquakes during SC23 and SC24, using the holography method. An extensive list of representation-learning-centered ML experiments are performed on this dataset, and two end-to-end models are analyzed from a solar-physics phenomenological standpoint. Figure \ref{fig:process.png} offers a step-by-step visualization of the process behind the proposed methodological setup. The following summary reiterates our most relevant findings and plans for future improvements.

\begin{itemize}
    \item Dataset construction: We created a comprehensive dataset of acoustic-emission power maps for ARs that contain sunquake signatures. These ML ready datasets are available in the linked repository. We emphasize that characteristics of this dataset include a large imbalance factor of $\approx$\,1 to 15\,--\,25 positive to negative ratio per dataset, very high noise patterns outside ARs, and the presence of artefacts of oscillatory moving features that result from the holographic analysis.

    \item Impact of custom augmentations: three domain-specific transforms were developed and introduced. They significantly improve our CL-based models performance on our difficult dataset: customized Random Erase, Solarized Low Pass Filter, Time Based Mix. 
    
    \item Dissimilar embeddings for sunquake start and end times: Both discussed CL models produce embeddings that have a high cosine distance towards their vicinity in the case of sunquake-transition frames, meaning that they are adept at predicting sunquake start and end frames. This is facilitated by the custom Time-Based-Mixing transform.

    \item Temporal and Spatial components: Our CL-trained models tend to predict a high number of apparently FP detections, compelling us to enhance temporal CL predictions with OD provided locations, to provide ML explainability. We found correlations between weaker acoustic-emission sources and solar eruptions accompanied by high-energy X-ray emission with spectral class $>\textrm{C}1.0$ in AIA and RHESSI data. Two such examples are discussed in depth, leading us to qualitatively hypothesize that our approach might be usable to detect weaker acoustic manifestations than previously possible. We stress that these two tentative detections could not be fully validated using helioseismology. A statistical meaningful analysis will be pursued in the future for a quantitative confirmation. 

    \item AutoEncoder approach limitations: No autoencoder-based approach, regardless of complexity, proved usable for our dataset. On a dataset by dataset basis, basic clusterization of sunquake-positive signals could be obtained for multiple sunquake individual datacubes.

    \item Custom augmentations as means of explainability: Although the custom Solarized Low Pass Filter augmentation significantly improves both CL models, we find it is not suitable for use a stand-alone tool for ML-output explainability.
    
    \item Unmet minimum sunquake duration criteria: The model described in Section \ref{ss-model3} is also sensible in predicting shorter duration  signals, be they acoustic in nature or not. This is explained by configuring the Time Based Mixing transform only on a series of three frames. In the future, we will experiment with increasing this window. This was not currently possible due to unfeasible computational costs.
    
    \item Unaddressed Spurious Correlations: False detections such as strong shifts in the gradient intensity in the ARs shadow, which are essentially noise, indicate that the models do not yet distinguish only true sunquake signatures. We believe a more fine-grained separation in signatures is needed, to encourage the models to produce sunquake representations that are dissimilar to their unfounded lookalikes. The additional representations might also prove to be of significant interest.
    
    \item Impact of Noise: Noise has long been the biggest enemy for astronomers and data scientists. While the ML techniques and custom transforms we introduced facilitate learning, they are not reliable for studying the effects of noise in their current state, as they all have a chance of obscuring the sunquake information. Moreover, conventional noise-reduction methods such as mean or Gaussian filters are also not reliable because noise patterns in egression-power maps are not distinctive enough from sunquake signatures. \\ \\ As noise continues to impose several challenges, additional pre-learning steps for performing noise reduction are paramount in improving the results and learning complexity. For this, we plan to repurpose the AutoEncoder model such that the respective model learns AR shapes, instead of sunquake signatures. This will also allow AutoEncoder training on additional datasets that do not necessarily contain sunquakes. Following this, a pre-learning step can possibly be applied to the input image to blacken out the area outside the AR shadow, obtained from the model's reconstruction, so that the latent space needed for the main CL training may be largely reduced.

\end{itemize}

\begin{acks}
The authors appreciate the very pertinent comments from the referee and editor, which were crucial towards improving this work. The authors thank Benoit Tremblay for the initial pre-submission review of this work. We acknowledge and appreciate the conversations and generous guidance provided by Alina Donea and Charles Lindsey towards pursuing this project. The authors also thank the Technical University of Cluj-Napoca for providing access to the DGX system which facilitated the execution of the presented experiments. 
Raw data and calibration instructions are obtained courtesy of NASA SDO/HMI, SOHO/MDI science teams. The authors welcome and appreciate the open data policy of the SDO, SOHO, and RHESSI missions. This research used version 4.0.X \citet{sunpy_community2020} of the SunPy open source software package. This work has made use of NASA's Astrophysics Data System (ADS). 
\end{acks}

\begin{fundinginformation}
A.R. Paraschiv and D.A. Lacatus are funded by the High Altitude Observatory of the National Center for Atmospheric Research, facilities that are sponsored by the National Science Foundation under cooperative agreement No. 1852977. D. Besliu-Ionescu was supported by a grant of the Romanian Ministry of Research, Innovation and Digitalization, CNCS/CCCDI-UEFISCDI, project number PN-III-P2-2.1-SOL-2021-2-0192, within PNCDI III. 
\end{fundinginformation}

\begin{dataavailability}
 The datasets generated and/or analyzed during the current study are available in our kaggle sunquakeNet repository, \href{www.doi.org/10.34740/kaggle/dsv/4111942}{\textsf{DOI: 10.34740/kaggle/dsv/4111942}}.
\end{dataavailability}

\begin{codeavailability}

The most current models can be provided by us upon reasonable request. 
\end{codeavailability}

\begin{ethics}
\begin{conflict}
The authors declare that they have no conflicts of interest.
\end{conflict}
\end{ethics}


 
\newpage
\bibliographystyle{spr-mp-sola}
\bibliography{bibliography}  

\IfFileExists{\jobname.bbl}{} {\typeout{}
\typeout{****************************************************}
\typeout{****************************************************}
\typeout{** Please run "bibtex \jobname" to obtain} \typeout{**
the bibliography and then re-run LaTeX} \typeout{** twice to fix
the references !}
\typeout{****************************************************}
\typeout{****************************************************}
\typeout{}}

\newpage
\appendix

\section{AutoEncoder approaches}
\label{sss-autoencoder}



As expected, MLPs, simple CNNs and Transfer Learning from ImageNet, which were first applied, failed in achieving good performance due to the complexity of the data, the highly imbalanced property and the transferred feature relevance.

For this reason, before attempting to apply CL and OD techniques described in this paper, we decided to use Representation Learning in the form of unsupervised, or self-supervised, AutoEncoder approaches. For these, a ResNet-6 \citep{He2016DeepRL} backbone is chosen, mainly because a more complex backbone would easily overfit given the small dataset size of SC23.

In self-supervised learning, original annotations are disregarded, but task-specific synthetic annotations are obtained from the data such that the learning is still a supervised one.

AutoEncoders are neural networks that reduce the dimensionality of the data and learn to reconstruct the data from the obtained encoding to a representation that is as similar to the original data as possible.

This category of neural networks was first introduced by \citet{HintonSalakhutdinov2006b} as an improvement to dimensionality reduction techniques such as Principal Component Analysis \citep{MACKIEWICZ1993303}. AutoEncoders were later proposed as generative models  \citep{Kingma2014aevb}, \citep{kingma2019introduction}, and further improved by \citet{Zhang2019PerceptualGA} by reducing training impediments given by inconsistencies between the data space and the latent space. We decided to choose this approach as it deals well with noisy data, noise being the main impediment for our holography datasets.

\subsection{Reduced AutoEncoder}
First, a plain AutoEncoder model with convolutional layers is implemented, with a latent space of size varying from 32 to 4096, with Mean Squared Error loss. The reconstructions for data samples are manually validated to confirm sunquake information is preserved. 
Due to varying morphological structure between events, reconstructions obtained from latent sizes lower than 1024 do not recover the originals, while larger latent sizes fail in boosting the classification performance even though they support good reconstructions.

\subsection{Variational AutoEncoder}
If the AutoEncoder objective is extended to include and simultaneously optimize both reconstruction and latent space distribution losses, the latent state for an observation is closer to others but deviating when necessary to describe distinctive features in the input \citep{kingma2019introduction}. 
This enables balanced latent-state representations of the input data, and it provides artificially generated reconstructions that follow the learned distributions. Following \citet{Kingma2014aevb}, we modify our model into a Variational AutoEncoder (VAE), by updating the Encoder to also output a distribution in the latent space with explicitly modeled variance.

Although VAE theoretically improves classifications through the learned representation, the experiments performed on our dataset still show weak reconstructions, depicting almost entirely black images. Classification results obtained from the latent representation of the input data also demonstrate that sunquake highlights are not captured in the latent space. During classification, experiments are performed using different loss functions, but even imbalance-specific losses, such as focal loss \citep{focalloss}, are still incapable of providing reliable results.

Additional different variants of VAE are utilized to find a more suitable design that  would i) provide reconstructions focused on the sunquake rather than the AR of the input sample, and that would; ii) be able to capture small variations around or inside the AR where sunquakes usually occur.  We take note of \citet{inbook}, who found that the typically used squared $L_2$-reconstruction loss function results in problems such as blurry images, and that it is also known to be sensitive to large noise. 

\citet{chen2019log} replace VAE's $L_2$-objective with the log hyperbolic cosine (log-cosh) loss, which behaves as $L_2$ at small values and as $L_1$ at large values. Compared to $L_2$, according to \citet{chen2019log}, the log-cosh loss improves the reconstruction without damaging the latent-space optimization, thus automatically keeping a balance between the reconstruction and the generation.

The images in our dataset are noisy, especially in areas outside ARs. Moreover, the sunquake signature represents only a small variation in the data for a short duration of time. Because of this, we believe that an objective that is less sensitive to noise could improve the model's focus.

\begin{figure}[t]
    \centering
    \begin{minipage}[t]{0.23\linewidth}
        \centering
        \includegraphics[width=1\textwidth]{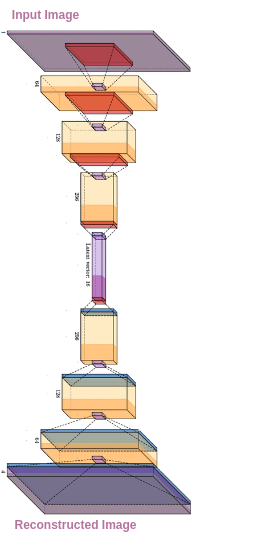}
    \end{minipage}
    \begin{minipage}[t]{0.76\linewidth}
        \centering
        \includegraphics[width=1\textwidth]{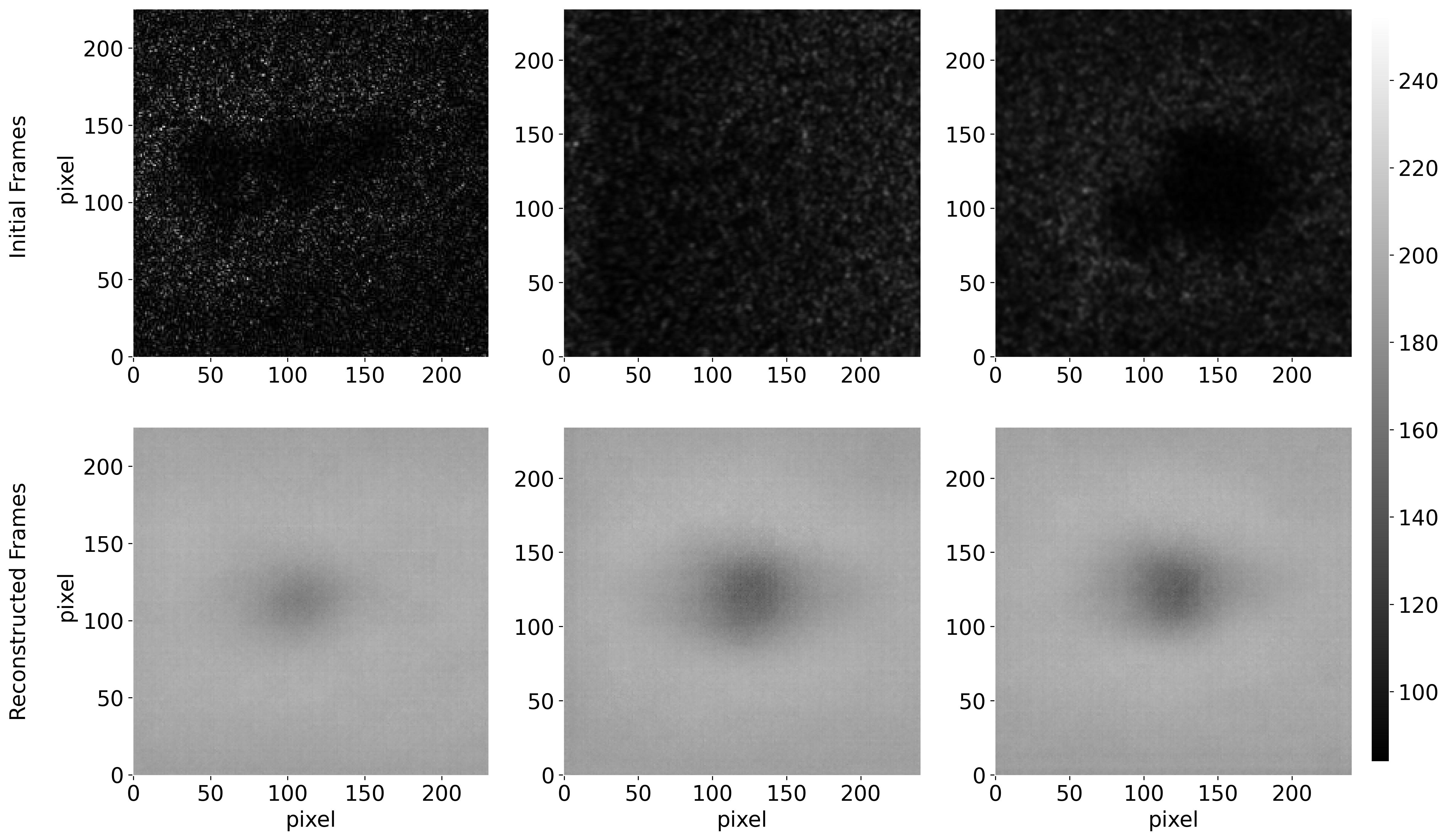}
        \vfill \vfill
    \end{minipage}
    \caption{AutoEncoder Architecture \citep{Xu_ae}\textit{(left)}, Variational AutoEncoder reconstructions of three randomly chosen 8-bit scaled intensity data samples from the ML prepared dataset \textit{(right)}.}
    \label{fig:recons.png}
\end{figure}

We find that the log-cosh loss benefits the training process and improves test accuracy by $\approx25\,\%$ when compared to the reduced AutoEncoder (see Table \ref{table:ae_results}).
Despite the improvement, sunquakes are still unidentifiable in the reconstructions with the naked eye. Reconstructions are blurry representations of the AR, with little to no detail, as depicted in Figure \ref{fig:recons.png}.

\subsection{Analysis of the Latent Features Quality}
To perform classification on top of the encoded data, various classifiers are trained over the encoding provided by the Log-cosh VAE: K-Nearest Neighbors (KNN) with and without bagging, Linear Classifier, Support Vector Classifier (SVC) \citep{cortes1995support}, Logistic Regression, fine-tuned unfrozen layers of the VAE, Stochastic Gradient Descent, and Random Forest \citep{ho1995random}. The results are not satisfactory, but are the first to provide a true meaningful input. A report of the best results achieved using Variational AutoEncoders as backbones for our models is displayed in Table \ref{table:ae_results}. Although still unreliable for classification, the reconstructions may prove potentially useful for tasks related to masking the AR with the purpose of reducing noise present in the dataset.

\begin{figure}[t]
    \centering
    \includegraphics[width=1\textwidth]{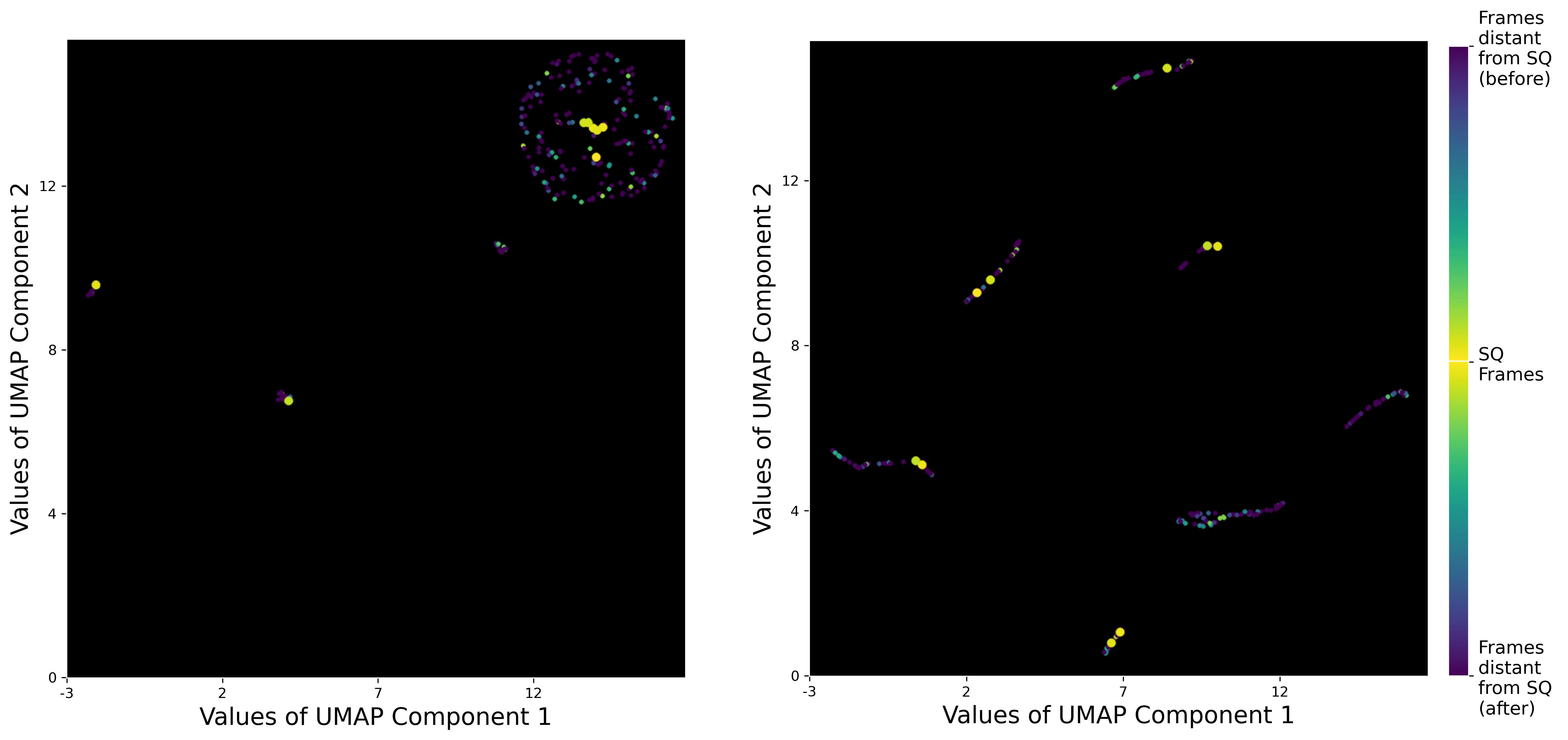}
    \caption{UMAP Components constructed from AutoEncoder-produced embeddings, for all samples in a singular event using 250 \textit{(left)} and 15 \textit{(right)} neighbors. Sample Count is 256. \textit{The colorbar goes symmetrically from yellow for frames corresponding to the sunquake, to purple for frames on either side of the sunquake.}}
    \label{fig:umap_ae.png}
\end{figure}

Figure \ref{fig:umap_ae.png} shows UMAPs built from the latent encoding of sunquake samples. In the left panel, we observe that clusters are formed in concentric circles, having the sunquake observations at the center. In the right panel, where fewer neighbors are selected, the clustering is lost. We infer that the observation’s background noise and the morphological structure of the AR play a significant role in the AutoEncoder output, as background counts in the AR are systematically lower, and morphologies are not consistent between the sunquake events. We found these effects to be significant enough to hinder a proper cluster-like classification of the sunquakes by using an AutoEncoder-based solution with reduced complexity.

It is important to underline that the clustering mentioned above is analyzed per sunquake event. 
The same analysis performed on the encoding extracted from the entire dataset shows no evident clusters. This supports our hypothesis that the encoding obtained using the mentioned unsupervised techniques is incapable of clustering data points from sunquakes with very different ARs in a two-dimensional setting. However, the few positive results combined with an increased precision (see Table \ref{table:ae_results}) when classifying the data show that some of the relevant aspects in the data are captured by the encoder, especially considering that no recurrent techniques are used in this approach.

We present the less-satisfactory performance of the AutoEncoder-based methods discussed above, in Table \ref{table:ae_results}.

\begin{table}[!t]
\tiny
 \setlength{\tabcolsep}{3.2pt}
\begin{tabular}{c c c c c c c c c }
 \hline
    
 Classifier &
    \multicolumn{2}{c}{K-NN (Bagging)} & 
        \multicolumn{2}{c}{SVC (Poly)} & 
            \multicolumn{2}{c}{Logistic Regression} & 
                \multicolumn{2}{c}{SGD} \\ 
 \hline
AE Type   & AE & Log-Cosh VAE & AE & Log-Cosh AE & AE & Log-Cosh VAE & AE & Log-Cosh VAE \\
 \hline

Precision   & 0.50 & 0.57 & 0.51 & \textbf{0.76} & 0.50 & 0.48 & 0.50 & 0.54  \\
Recall      & 0.46 & 0.56 & 0.52 & \textbf{0.59} & 0.54 & 0.48 & 0.50 & 0.55  \\
F1-Score    & 0.39 & 0.56 & 0.05 & \textbf{0.61} & 0.39 & 0.48 & 0.39 & 0.54  \\
Accuracy    & 0.62 & 0.82 & 0.05 & \textbf{0.88} & 0.59 & 0.78 & 0.61 & 0.74  \\ [1ex]
 \hline
 Metric Avg. & 0.49 & 0.62 & 0.28 & \textbf{0.71} & 0.5 & 0.55 & 0.5 & 0.59 \\ [1ex]
 \hline
\end{tabular}
\caption{Macro Average performances of different classifiers over AutoEncoder-produced embeddings, trained with SMOTE augmentation, on the SC23 test data (171 negative and 26 positive samples)}
\label{table:ae_results}
\centering
\end{table}

\section{Self-supervised and Supervised Contrastive Learning approaches}

This section describes the experiments performed using self-supervised and supervised contrastive learning approaches that led to establishing the two-step learning approach described in Section \ref{sss-contrastive}.

\subsection{Self-Supervised Contrastive Learning}
\label{sss-self-sup-cl}
The framework proposed by \citet{pmlr-v119-chen20j} is followed, where, for all the images in each data batch, two augmentations are drawn out so that a double-sized batch is obtained. Specific details on what augmentations were used in our experiments are presented in Subsection \ref{sss-augmentations}. Batch data are then passed through the CL model, where SimCLR loss \citep{pmlr-v119-chen20j} is used. The loss is defined as:
\begin{equation}
l_{i,j} = -\textrm{log} \frac{\textrm{e}^\frac{\textrm{sim}(z_{i} z_{j})}{\tau}}{\sum_{k=1}^{2N} \textrm{mask}_{[k\neq i]}\textrm{e}^\frac{\textrm{sim}(z_{i} z_{k})}{\tau}}.
\label{eq1}
\end{equation}

In the equation above, $(i,j)$ represents a positive pair of samples, $\textrm{mask}$ is an indicator function evaluating to $1$ if $k \neq i $, $z$ represents the embedding obtained from passing the input through the model, $\textrm{sim}$ represents cosine similarity. $\tau$ is a parameter used to regularize the contribution of representations \citep{pmlr-v119-chen20j}.

At low temperatures, smaller distances are favored, which leads to better capturing of minor differences but makes training more difficult since numerical instability increases \citep{supcon}.
In the experiments performed on our data, a temperature factor lower than $\tau = 0.07$ is not usable due to this instability, and generally values of $\tau \in [0.07, 0.1]$ work best.

We experiment with dimension size for the encoding (latent representation resulting from the backbone) and the projection (latent representation resulting from the contrastive head) varying from 20 and up to 2048 features.

Because the discussed approach is self-supervised, labels are not considered. The classification is performed using a weighted Cross Entropy loss after the contrastive training is finalized. We experiment with freezing network layers in the backbone network during classification. This process prevents any updates to the weights of the frozen layers of the network and assures that changes inferring from classification are not so drastic as to erase relevant information learned by the network during CL. 

When classifying the reduced representations, we experiment with using both the encoding and the projection. On average, results of this approach are similar to those of the AutoEncoder methods. Seeing that this type of approach is not able to capture sunquake signatures in the latent representation, we shift to the recently emerging Supervised CL \citep{supcon}. 

\subsection{Supervised Contrastive Learning}
\label{sss-sup-cl}

The implementation of a supervised-contrastive model as compared to the self-supervised one differs only in  loss function. The Supervised Contrastive loss function \citep{supcon}, written in the same manner as Equation \ref{eq-supcon} is defined for each anchor $i$ as:

\begin{equation}
l_{i} = \sum_{i \in I} \frac{-1}{|P(i)|} \sum_{p \in P(i)} \textrm{log} \frac{\textrm{e}^\frac{\textrm{sim}(z_{i} z_{p})}{\tau}}{\sum_{k=1}^{2N} \textrm{mask}_{[k\neq i]}e^\frac{\textrm{sim}(z_{i} z_{k})}{\tau}}. \label{eq-supcon}
\end{equation} 

The meaning of most parameters is maintained as in Equation \ref{eq1}. $P$ represents the set of indices of the positives, which are all of the samples that share the same class as the anchor and their augmentations.

To verify the validity of the implementation of our models, an experiment is performed on a commonly used small-scale ML dataset, \textsf{CIFAR-10} \citep{Krizhevsky09learningmultiple}. The model setup is kept in place, and for augmentations both \textsf{CIFAR} specific augmentations and ours are used. Findings of this experiment show that the model is only capable of convergence after $\approx$100 epochs (top-1 classification accuracy $>$ 90.0). Up to this point, the model is stuck in a local minimum. We deduce that several more epochs are needed to improve convergence on our data, especially because the data structure has a higher complexity, a larger imbalance factor, is noisier, and is larger in size.

Increasing the number of epochs for the supervised model shows a similar behavior in terms of local minima, but the class imbalance factor still posed a great impact on the classification results. A Two-step approach combining self-supervised with supervised CL models was found to be robust in capturing sunquake features, as shown in Section \ref{sss-contrastive}.

\section{Data augmentation and sampling approaches}
\label{sss-augmentations2}
This section describes all the data augmentation methods briefly mentioned in Section \ref{sss-augmentations}, following the categories proposed by  \citep{augmentations2022Yang}.

\subsection{Geometry}
\textit{Flips:} We apply horizontal and vertical flips as a method of upsampling by adding flipped copies of the positive samples to the dataset. We also apply flip operations at runtime, to ensure that the transforms do not infer external information to the model regarding positive samples.

\textit{Crops:} The random-crop transform is typically used in most CL approaches. We found it unsuitable for our data since the relevant area for the sunquake signature is very small, and there is a chance of cropping out that specific area. 
\citet{augmentations2022Yang} also argue that this might not be a label-preserving transformation. Adding this augmentation to our data at runtime prevents convergence of supervised methods.

\textit{Rotations:} We apply right rotation on axes of $90$\textdegree, $180$\textdegree, and $270$\textdegree. Because our data and labels are invariant to rotations, this transformation is usable. As with flip operations, rotations are used as a means to upsample our dataset with the purpose of reducing class imbalance. To make sure that the model does not learn to associate geometric transforms with a positive class, we apply at least the same number of rotated frames to the negative samples at runtime.

\textit{Translations:} Because the data are not translation invariant, we avoid this category of transforms to preserve the labels.\\

\subsection{Kernel Filters}
\textit{Gaussian Blur:} We experiment with Gaussian Blur and argue that it is not always label preserving, hindering the sunquake information for some of the positive samples. We found it to not provide notable improvements in the learning process, and thus we discard this transformation. \\
    
\subsection{Mixing Images}
A custom \textit{Custom Time-Based Mixing} augmentation is described in Section \ref{sss-augmentations}.
    
\subsection{Auto Augment Policies}
\textit{ImageNet and CIFAR Auto Augment Policies:} During the experiment with the CIFAR dataset described in Section \ref{sss-contrastive} we apply the Auto Augment policies generated for ImageNet and \textsf{CIFAR} on both the \textsf{CIFAR-10} dataset and ours, and while the ImageNet policy works well on \textsf{CIFAR-10}, when applied to our dataset the results were found to be poor.
    
\subsection{Color Space} 
\label{ssss-augmentations2-colorspace}
As our data is gray-scale, we do not use many color altering transforms (i.e. color jitter). We experiment with contrast, posterize, brightness, and solarization and manually review the augmented data to analyze the impact on the sunquake preservation. With the exception below, color-space transforms are ineffective.

We applied a color space like transform, \textit{Custom Solarized Low Pass Filter}, as described in Section \ref{sss-augmentations}.

\begin{figure}[t]
\centering
\includegraphics[width=\textwidth]{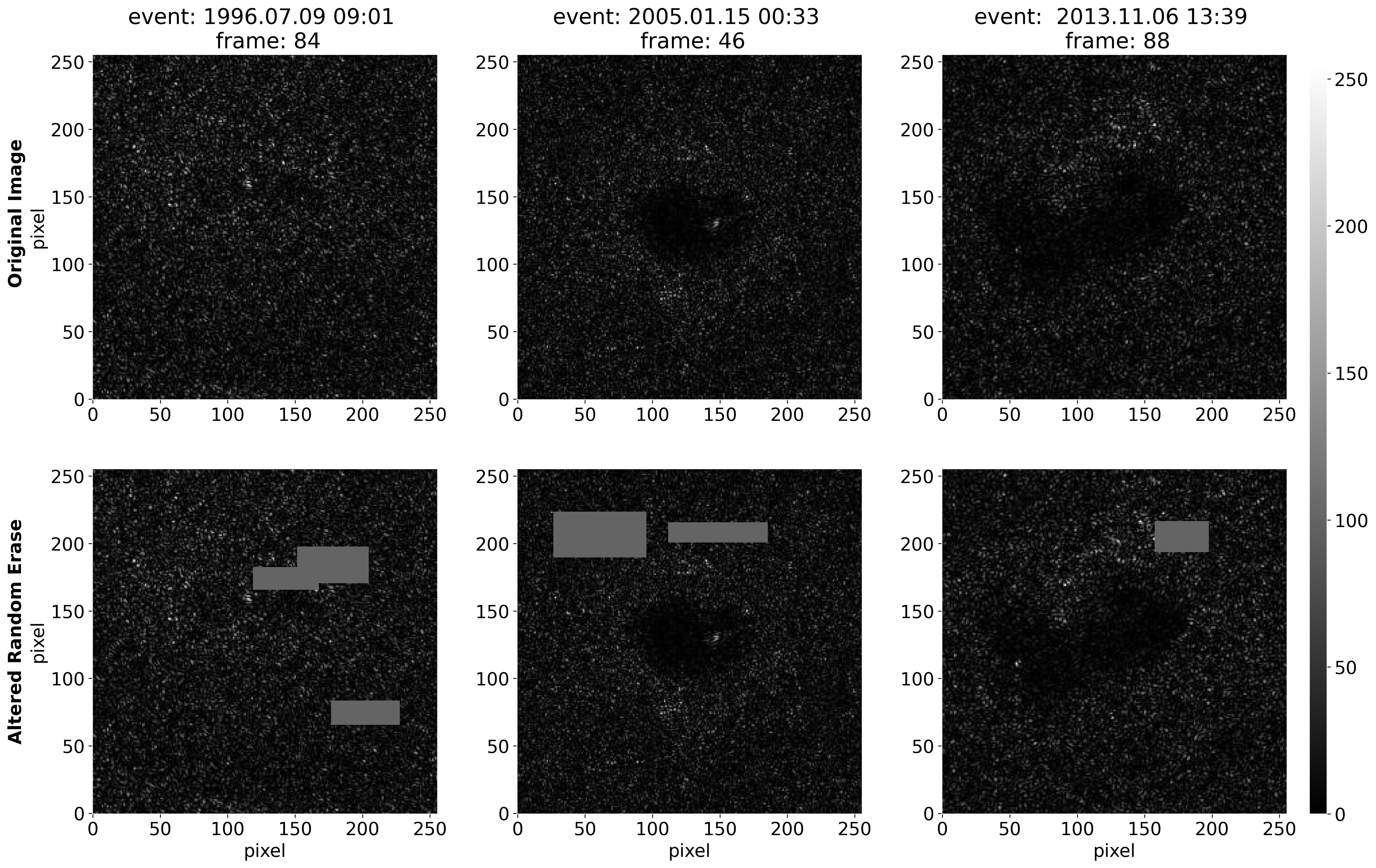}
\caption{Altered Random Erase applied to a selection of 8-bit intensity scaled frames from the SC23 and SC24 dataset. Frame indices and events are mentioned in the column titles. }
\label{fig:altered_random_erase.png}
\end{figure}

\subsection{Occlusion}
\label{ssss-augmentations2-occlusion}
\textit{Random Erasing:} This type of augmentation developed by \citet{Zhong2020RandomED} and inspired by the mechanisms of dropout regularization \citep{augmentations2022Yang} is typically used for scenarios where the detection mechanism tends to fail due to occlusions. It forces the model to learn more relevant features about an image, while preventing overfitting. In typical CL tasks, occlusion is achieved via the Random Crop transform. In this case, we aim to shift the model's focus from the AR to the sunquake.  In Section \ref{sss-augmentations} we describe customizing this type of augmentation to force the network to pay attention to all areas inside the image. We do this by occluding random areas of both the noisy regions and the AR (see Figure \ref{fig:altered_random_erase.png}).

\section{IoU Metric Limitations in Sunquake Detection}
\label{od-appendix}

\begin{figure}[t]
\centering
\includegraphics[width=\textwidth]{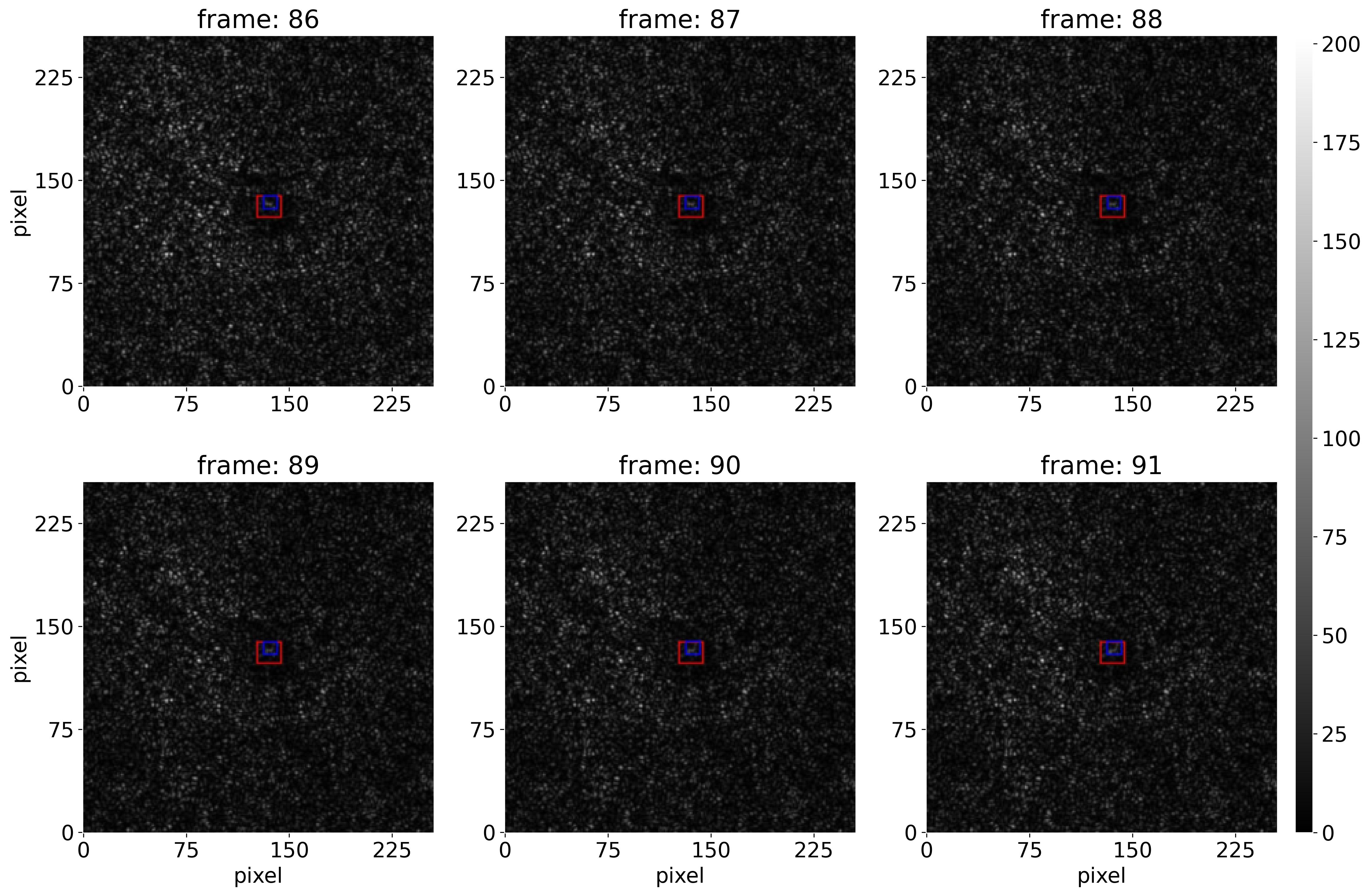}
\caption{Position of identified OD kernel marked by the 
\textit{blue box} vs. GT kernels marked by the \textit{red box} for six 8-bit scaled intensity data samples from the event 07 September 2011 22:32 in SC24. IoU varies between 30\,\% and 35\,\% for the given frames.}
\label{fig:od_c24.png}
\end{figure}

Figure \ref{fig:od_c24.png} highlights the OD prediction boxes and the GT boxes for an event in SC24, where the discrepancy between a correct detection and a low IoU discussed in Section \ref{sss-objectdet} is exemplified. For this event, despite a valid signature identification, the average OD IoU for all frames is small (22.3\,\%), and individual IoUs for the frames depicted in the figure are as follows:  34.1\,\%, 32.7\,\%, 30.8\,\%, 33\,\%, 34.2\,\%, and 36\,\%.

\newpage
\section{List of Acronyms}
\[
\begin{array}{ll}

{SQ}  & {Sunquake}\\
{SC}  & {Solar\ Cycle}\\
{AR}  & {Active\ Region}\\
{SDO}  & {Solar\ Dynamics\ Observatory}\\
{SOHO}  & {Solar\ and\ Heliospheric\ Observatory}\\
{HMI}  & {Helioseismic\ and\ Magnetic\ Imager}\\
{MDI}  & {Solar\ and\ Heliospheric\ Observator}\\
{AIA}  & {Atmospheric\ Imaging\ Assembly}\\
{RHESSI} & {Ramaty\ High\ Energy\ Solar\ Spectroscopic\ Image}\\
{ML}  & {Machine\ Learning}\\
{MLP}  & {Multilayer\ Perceptron}\\
{CNN}  & {Convolutional\ Neural\ Networks}\\
{R-CNN}  & {Region\,-\,Based\ Convolutional\ Neural\ Networks}\\
{VAE}  & {Variational\ AutoEncoder}\\
{CL}  & {Contrastive\ Learning}\\
{OD}  & {Object\ Detection}\\
{SVC}  & {Support\ Vector\ Classifier}\\
{SGD}  & {Stochastic\ Gradient\ Descent}\\
{RBF}  & {Radial\ basis\ function}\\
{PCA}  & {Principal\ Component\ Analysis}\\
{UMAP}  & {Uniform\ Manifold\ Approximation\ and\ Projection}\\
{SMOTE}  & {Synthetic\ Minority\ Over\,-\,Sampling\ Technique}\\
{FP}  & {False\ Positive}\\
{FN}  & {False\ Negative}\\
{TP}  & {True\ Positive} \\
{TN}  & {True\ Negative} \\
{GT}  & {Ground\ Truth}\\
{IoU} & {Intersection\ over\ Union}\\ 
\end{array}
\]

\end{article} 

\end{document}